\input harvmac

\newdimen\tableauside\tableauside=1.0ex
\newdimen\tableaurule\tableaurule=0.4pt
\newdimen\tableaustep
\def\phantomhrule#1{\hbox{\vbox to0pt{\hrule height\tableaurule width#1\vss}}}
\def\phantomvrule#1{\vbox{\hbox to0pt{\vrule width\tableaurule height#1\hss}}}
\def\sqr{\vbox{%
  \phantomhrule\tableaustep
  \hbox{\phantomvrule\tableaustep\kern\tableaustep\phantomvrule\tableaustep}%
  \hbox{\vbox{\phantomhrule\tableauside}\kern-\tableaurule}}}
\def\squares#1{\hbox{\count0=#1\noindent\loop\sqr
  \advance\count0 by-1 \ifnum\count0>0\repeat}}
\def\tableau#1{\vcenter{\offinterlineskip
  \tableaustep=\tableauside\advance\tableaustep by-\tableaurule
  \kern\normallineskip\hbox
    {\kern\normallineskip\vbox
      {\gettableau#1 0 }%
     \kern\normallineskip\kern\tableaurule}%
  \kern\normallineskip\kern\tableaurule}}
\def\gettableau#1 {\ifnum#1=0\let\next=\null\else
  \squares{#1}\let\next=\gettableau\fi\next}

\tableauside=1.0ex
\tableaurule=0.4pt
\input epsf
\noblackbox

\def\l{{\lambda}}



\def\unlockat{\catcode`\@=11}
\def\lockat{\catcode`\@=12}

\unlockat

\def\newsec#1{\global\advance\secno by1\message{(\the\secno. #1)}
\global\subsecno=0\global\subsubsecno=0\eqnres@t\noindent
{\bf\the\secno. #1}
\writetoca{{\secsym} {#1}}\par\nobreak\medskip\nobreak}
\global\newcount\subsecno \global\subsecno=0
\def\subsec#1{\global\advance\subsecno
by1\message{(\secsym\the\subsecno. #1)}
\ifnum\lastpenalty>9000\else\bigbreak\fi\global\subsubsecno=0
\noindent{\it\secsym\the\subsecno. #1}
\writetoca{\string\quad {\secsym\the\subsecno.} {#1}}
\par\nobreak\medskip\nobreak}
\global\newcount\subsubsecno \global\subsubsecno=0
\def\subsubsec#1{\global\advance\subsubsecno
\message{(\secsym\the\subsecno.\the\subsubsecno. #1)}
\ifnum\lastpenalty>9000\else\bigbreak\fi
\noindent\quad{\secsym\the\subsecno.\the\subsubsecno.}{#1}
\writetoca{\string\qquad{\secsym\the\subsecno.\the\subsubsecno.}{#1}}
\par\nobreak\medskip\nobreak}

\def\subsubseclab#1{\DefWarn#1\xdef
#1{\noexpand\hyperref{}{subsubsection}%
{\secsym\the\subsecno.\the\subsubsecno}%
{\secsym\the\subsecno.\the\subsubsecno}}%
\writedef{#1\leftbracket#1}\wrlabeL{#1=#1}}
\lockat

\def\IL{{\relax{\rm I\kern-.18em L}}}
\def\IH{{\relax{\rm I\kern-.18em H}}}
\def\IR{{\relax{\rm I\kern-.18em R}}}
\def\IE{{\relax{\rm I\kern-.18em E}}}
\def\IC{{\relax\hbox{$\inbar\kern-.3em{\rm C}$}}}
\def\IZ{{\relax\ifmmode\mathchoice
{\hbox{\cmss Z\kern-.4em Z}}{\hbox{\cmss Z\kern-.4em Z}}
{\lower.9pt\hbox{\cmsss Z\kern-.4em Z}}
{\lower1.2pt\hbox{\cmsss Z\kern-.4em Z}}\else{\cmss Z\kern-.4em
Z}\fi}}

\def\CF {{\cal F}}

\def\CL {{\cal L}}

\def\CO {{\cal O}}

\def\CG {{\cal G}}

\def\CW {{\cal W}}


\def\CO {{\cal O}}

\font\manual=manfnt \def\dbend{\lower3.5pt\hbox{\manual\char127}}

\def\IZ{{\relax\ifmmode\mathchoice
{\hbox{\cmss Z\kern-.4em Z}}{\hbox{\cmss Z\kern-.4em Z}}
{\lower.9pt\hbox{\cmsss Z\kern-.4em Z}}
{\lower1.2pt\hbox{\cmsss Z\kern-.4em Z}}\else{\cmss Z\kern-.4em
Z}\fi}}

\def\CW {{\cal W}}
\def\CL {{\cal L}}

\def\CO {{\cal O}}

\def\dim{{\hbox{dim}}}


\def\IZ{{\relax\ifmmode\mathchoice
{\hbox{\cmss Z\kern-.4em Z}}{\hbox{\cmss Z\kern-.4em Z}}
{\lower.9pt\hbox{\cmsss Z\kern-.4em Z}}
{\lower1.2pt\hbox{\cmsss Z\kern-.4em Z}}\else{\cmss Z\kern-.4em
Z}\fi}}
\def\IB{{\relax{\rm I\kern-.18em B}}}
\def\IC{{\relax\hbox{$\inbar\kern-.3em{\rm C}$}}}
\def\ID{{\relax{\rm I\kern-.18em D}}}
\def\IE{{\relax{\rm I\kern-.18em E}}}
\def\IF{{\relax{\rm I\kern-.18em F}}}
\def\IG{{\relax\hbox{$\inbar\kern-.3em{\rm G}$}}}
\def\IGa{{\relax\hbox{${\rm I}\kern-.18em\Gamma$}}}
\def\IH{{\relax{\rm I\kern-.18em H}}}
\def\II{{\relax{\rm I\kern-.18em I}}}
\def\IK{{\relax{\rm I\kern-.18em K}}}
\def\IP{{\relax{\rm I\kern-.18em P}}}

\def\l{{\lambda}}

\def\inbar{\,\vrule height1.5ex width.4pt depth0pt}

\font\cmss=cmss10 \font\cmsss=cmss10 at 7pt
\def\IR{\relax{\rm I\kern-.18em R}}
\def\IT{\relax{\rm I\kern-.45em T}}

\def\Tr{{\rm Tr}}


\def\boxit#1{\vbox{\hrule\hbox{\vrule\kern8pt
\vbox{\hbox{\kern8pt}\hbox{\vbox{#1}}\hbox{\kern8pt}}
\kern8pt\vrule}\hrule}}
\def\mathboxit#1{\vbox{\hrule\hbox{\vrule\kern8pt\vbox{\kern8pt
\hbox{$\displaystyle #1$}\kern8pt}\kern8pt\vrule}\hrule}}


\def\inbar{\,\vrule height1.5ex width.4pt depth0pt}

\font\cmss=cmss10 \font\cmsss=cmss10 at 7pt
\def\IR{\relax{\rm I\kern-.18em R}}

\def\Tr{{\rm Tr}}


\def\exp{{\rm exp}}

\let\includefigures=\iftrue
\newfam\black
\includefigures

\input epsf
\def\plb#1 #2 {Phys. Lett. {\bf B#1} #2 }
\long\def\del#1\enddel{}
\long\def\new#1\endnew{{\bf #1}}
\let\<\langle \let\>\rangle

\def\figin{\epsfcheck\figin}\def\figins{\epsfcheck\figins}
\def\epsfcheck{\ifx\epsfbox\UnDeFiNeD
\message{(NO epsf.tex, FIGURES WILL BE IGNORED)}
\gdef\figin##1{\vskip2in}\gdef\figins##1{\hskip.5in} blank space instead
\else\message{(FIGURES WILL BE INCLUDED)}
\gdef\figin##1{##1}\gdef\figins##1{##1}\fi}
\def\DefWarn#1{}
\def\figinsert{\goodbreak\midinsert}
\def\ifig#1#2#3{\DefWarn#1\xdef#1{fig.~\the\figno}
\writedef{#1\leftbracket fig.\noexpand~\the\figno}
\figinsert\figin{\centerline{#3}}\medskip
\centerline{\vbox{\baselineskip12pt
\advance\hsize by -1truein\noindent
\footnotefont{\bf Fig.~\the\figno:} #2}}
\bigskip\endinsert\global\advance\figno by1}
\else
\def\ifig#1#2#3{\xdef#1{fig.~\the\figno}
\writedef{#1\leftbracket fig.\noexpand~\the\figno}
\figinsert\figin{\centerline{#3}}\medskip
\centerline{\vbox{\baselineskip12pt
\advance\hsize by -1truein\noindent
\footnotefont{\bf Fig.~\the\figno:} #2}}
\bigskip\endinsert
\global\advance\figno by1}
\fi

\input xy
\xyoption{all}
\font\cmss=cmss10 \font\cmsss=cmss10 at 7pt
\def\inbar{\,\vrule height1.5ex width.4pt depth0pt}
\def\IC{{\relax\hbox{$\inbar\kern-.3em{\rm C}$}}}
\def\IP{{\relax{\rm I\kern-.18em P}}}
\def\IF{{\relax{\rm I\kern-.18em F}}}
\def\IZ{\relax\ifmmode\mathchoice
{\hbox{\cmss Z\kern-.4em Z}}{\hbox{\cmss Z\kern-.4em Z}}
{\lower.9pt\hbox{\cmsss Z\kern-.4em Z}}
{\lower1.2pt\hbox{\cmsss Z\kern-.4em Z}}\else{\cmss Z\kern-.4em
Z}\fi}
\def\IR{{\relax{\rm I\kern-.18em R}}}
\def\IQ{\relax\hbox{\kern.25em$\inbar\kern-.3em{\rm Q}$}}

\def\pmb#1{\setbox0=\hbox{#1}%
 \kern-.025em\copy0\kern-\wd0
 \kern.05em\copy0\kern-\wd0
 \kern-.025em\raise.0433em\box0 }
\font\cmss=cmss10
\font\cmsss=cmss10 at 7pt
\def\rlx{\relax\leavevmode}
\def\Cop{\relax\,\hbox{$\inbar\kern-.3em{\rm C}$}}
\def\Rop{\relax{\rm I\kern-.18em R}}
\def\Nop{\relax{\rm I\kern-.18em N}}
\def\Pop{\relax{\rm I\kern-.18em P}}
\def\Zop{\rlx\leavevmode\ifmmode\mathchoice{\hbox{\cmss Z\kern-.4em Z}}
 {\hbox{\cmss Z\kern-.4em Z}}{\lower.9pt\hbox{\cmsss Z\kern-.36em Z}}
 {\lower1.2pt\hbox{\cmsss Z\kern-.36em Z}}\else{\cmss Z\kern-.4em
 Z}\fi}

\def\inbar{\,\vrule height1.5ex width.4pt depth0pt}
\def\IC{{\relax\hbox{$\inbar\kern-.3em{\rm C}$}}}
\def\IP{{\relax{\rm I\kern-.18em P}}}
\def\IF{{\relax{\rm I\kern-.18em F}}}
\def\IZ{\relax\ifmmode\mathchoice
{\hbox{\cmss Z\kern-.4em Z}}{\hbox{\cmss Z\kern-.4em Z}}
{\lower.9pt\hbox{\cmsss Z\kern-.4em Z}}
{\lower1.2pt\hbox{\cmsss Z\kern-.4em Z}}\else{\cmss Z\kern-.4em
Z}\fi}
\def\IR{{\relax{\rm I\kern-.18em R}}}
\def\IT{{\mathchoice {\setbox0=\hbox{$\displaystyle\rm
T$}\hbox{\hbox to0pt{\kern0.3\wd0\vrule height0.9\ht0\hss}\box0}}
{\setbox0=\hbox{$\textstyle\rm T$}\hbox{\hbox
to0pt{\kern0.3\wd0\vrule height0.9\ht0\hss}\box0}}
{\setbox0=\hbox{$\scriptstyle\rm T$}\hbox{\hbox
to0pt{\kern0.3\wd0\vrule height0.9\ht0\hss}\box0}}
{\setbox0=\hbox{$\scriptscriptstyle\rm T$}\hbox{\hbox
to0pt{\kern0.3\wd0\vrule height0.9\ht0\hss}\box0}}}}
\def\bbbti{{\mathchoice {\setbox0=\hbox{$\displaystyle\rm
T$}\hbox{\hbox to0pt{\kern0.3\wd0\vrule height0.9\ht0\hss}\box0}}
{\setbox0=\hbox{$\textstyle\rm T$}\hbox{\hbox
to0pt{\kern0.3\wd0\vrule height0.9\ht0\hss}\box0}}
{\setbox0=\hbox{$\scriptstyle\rm T$}\hbox{\hbox
to0pt{\kern0.3\wd0\vrule height0.9\ht0\hss}\box0}}
{\setbox0=\hbox{$\scriptscriptstyle\rm T$}\hbox{\hbox
to0pt{\kern0.3\wd0\vrule height0.9\ht0\hss}\box0}}}}

\def\bX{{\overline X}}

\def\sevenpoint{
  \font\seveni=cmmi7
  \font\seven=cmsy7
  \font\sevensy=cmsy7
  \font\fivei=cmmi5
  \font\fivesy=cmsy5
  \font\it=cmti7
  \font\bf=cmbx7
  \font\sl=cmsl8
  \font\sixrm=cmr6
  \font\sixi=cmmi6
  \font\sixsy=cmsy6
  \textfont0= \sevenrm \scriptfont0=\sixrm 
\scriptscriptfont0=\fiverm
  \def\rm{\fam0 \sevenrm}
  \textfont1=\seveni  \scriptfont1=\sixi 
\scriptscriptfont1=\fivei
  \def\mit{\fam1 } \def\oldstyle{\fam1 \seveni}
  \textfont2=\sevensy \scriptfont2=\sixsy 
\scriptscriptfont2=\fivesy
\def\doublespace{\baselineskip=18pt\lineskip=0pt
\lineskiplimit=-5pt}
\def\singlespace{\baselineskip=9pt\lineskip=0pt
\lineskiplimit=-5pt}
\def\oneandahalf{\baselineskip=15pt\lineskip=0pt
\lineskiplimit=-5pt}
}

\nref\AAHV{B. Acharya, M. Aganagic, K. Hori and C. Vafa, ``Orientifolds, Mirror
Symmetry and Superpotentials," hep-th/0202208.}
\nref\AKMV{M. Aganagic, A. Klemm, M. Mari\~no and  C. Vafa, ``The Topological Vertex," hep-th/0305132.}
\nref\AKV{M. Aganagic, A. Klemm, and  C. Vafa, ``Disk Instantons, Mirror Symmetry and
the Duality Web,''Z.\ Naturforsch.\ A {\bf 57}, 1 (2002), hep-th/0105045.}
\nref\AMV{M. Aganagic, M. Mari\~no and C. Vafa, ``All Loop Topological Strings
Amplitudes from Chern-Simons Theory," hep-th/0206164.}
\nref\AV{M.~Aganagic and C.~Vafa, 
``$G_2$ manifolds, Mirror Symmetry and Geometric Engineering,'' hep-th/0110171.}
\nref\AGNT{I. Antoniadis, E.~Gava, K.~S.~Narain and T.~R.~Taylor,
``Topological Amplitudes in String Theory,''
Nucl.\ Phys.\ B {\bf 413}, 162 (1994), hep-th/9307158.}
\nref\BCOV{M.~Bershadsky, S.~Cecotti, H.~Ooguri and C.~Vafa,
``Kodaira-Spencer Theory of Gravity and Exact Results for Quantum String
Amplitudes,''
Commun.\ Math.\ Phys.\  {\bf 165}, 311 (1994), hep-th/9309140.}
\nref\BH{I. Brunner and K. Hori, ``Orientifolds and Mirror Symmetry," hep-th/0303135. }
\nref\CKYZ{T.-M. Chiang, A. Klemm, S.-T. Yau and E. Zaslow, ``Local Mirror Symmetry:
Calculations and Interpretations," ATMP {\bf 3} (1999) 495-565, hep-th/9903053.}
\nref\CVI{P. Cvitanovic, {\it Group Theory}, {{\tt http://www.cns.gatech.edu/predrag.}}}
\nref\DFGi{D.-E. Diaconescu, B. Florea and A. Grassi, ``Geometric
Transitions and Open String Instantons," ATMP {\bf 6} (2003) 619-642, hep-th/0205234.}
\nref\DFGii{D.-E. Diaconescu, B. Florea and A. Grassi, ``Geometric Transitions,
del Pezzo Surfaces and Open String Instantons," ATMP {\bf 6} (2003) 643, hep-th/0206163.}
\nref\DFM{D.-E. Diaconescu, B. Florea and A. Misra, ``Orientifolds, Unoriented
Instantons and Localization," JHEP {\bf 0307}, 041 (2003), hep-th/0305021.}
\nref\EK{T.~Eguchi and H.~Kanno, ``Geometric Transitions, Chern-Simons Gauge
Theory and Veneziano Type Amplitudes,'' Phys.\ Lett.\ B {\bf 585}, 163 (2004),
hep-th/0312234.}
\nref\FH{W. Fulton and J. Harris, {\it Representation Theory: A First Course}, Springer Verlag, 1997, 568 p.}
\nref\GP{T. Graber and R. Pandharipande, ``Localization of Virtual Classes,'' Invent. Math. {\bf 135}, 487 (1999) 
math.AG/9708001.}
\nref\GV{R. Gopakumar and C. Vafa, ``On the Gauge Theory/Geometry Correspondence," ATMP {\bf 3} (1999) 1415, hep-th/9811131.}
\nref\GViii{R. Gopakumar and C. Vafa, ``M-Theory and Topological Strings -II," hep-th/9812127.}
\nref\guada{E. Guadaganini, ``The Universal Link Polynomial,''
Int.\ J.\ Mod.\ Phys.\ A {\bf 7}, 877 (1992).}
\nref\Hal{K. Hori et al, {\it Mirror Symmetry}, C. Vafa and E. Zaslow, eds.,
Clay Mathematics Monographs, American Mathematical Society, 2003, 929p.}
\nref\king{R.C. King, ``Modification Rules and Products of
Irreducible Representations of the Unitary, Orthogonal, and Symplectic Groups,''
J. Math. Phys. {\bf 12}, 1588 (1971).}
\nref\Ko{M. Kontsevich, ``Enumeration of Rational Curves via Torus Actions," hep-th/9405035.}
\nref\Li{D. E. Littlewood, {\it The Theory of Group Characters and Matrix Representations of Groups}, 
Oxford, The Clarendon Press, 1940, 292 p.}
\nref\Macdonald{I.G. Macdonald, {\it Symmetric functions and Hall polynomials},
Oxford University Press, 1995.}
\nref\MVi{M. Mari\~no and C. Vafa, ``Framed Knots at Large $N$," hep-th/0108064.}
\nref\ORV{A.~Okounkov, N.~Reshetikhin and C.~Vafa, ``Quantum Calabi-Yau
and Classical Crystals,'' hep-th/0309208.}
\nref\OV{H. Ooguri and C. Vafa, ``Knot Invariants and Topological Strings," Nucl. Phys. B {\bf 577} 
(2000) 419-438, hep-th/9912123.}
\nref\Ra{S. Ramgoolam, ``Comment on Two Dimensional $O(N)$ and $Sp(N)$ Yang Mills
Theories as String Theories," Nucl. Phys. B {\bf 418} (1994) 30-44, hep-th/9307085.}
\nref\SV{S. Sinha and C. Vafa, ``$SO$ and $Sp$ Chern-Simons at Large $N$," hep-th/0012136.}
\nref\Wii{E. Witten, ``Chern-Simons Gauge Theory as a String Theory," in {\it The Floer Memorial Volume}, H. Hofer, C. H. Taubes, 
A. Weinstein and E. Zehner, eds., Birkh\"auser 1995, p.637, hep-th/9207094.}
\nref\Wiii{E. Witten, ``Quantum Field Theory and the Jones Polynomial," Commun. Math.
Phys. {\bf 121} (1989) 351.}
\Title{
\vbox{
\baselineskip12pt
\hbox{CERN-PH-TH/2004-077}
\hbox{hep-th/0405083}}}
{\vbox{\vskip 19pt
\vbox{\centerline{Counting Higher Genus Curves with Crosscaps}
\vskip.1cm
\centerline{in Calabi-Yau Orientifolds}
}}}
\vskip 15pt
\centerline{Vincent Bouchard\footnote{$^\natural$}
{{{\tt bouchard@maths.ox.ac.uk}}}, Bogdan Florea\footnote{$^\sharp$}
{{{\tt florea@physics.rutgers.edu}}} and Marcos
Mari\~no\footnote{$^\flat$}{{{\tt marcos@mail.cern.ch}; also at Departamento de
Matem\'atica, IST, Lisboa, Portugal.}}}
\bigskip
\medskip
\centerline{$^\natural${\it Mathematical Institute, University of Oxford,}}
\centerline{\it{24-29 St. Giles', Oxford OX1 3LB, England}}
\centerline{$^\sharp${\it Department of Physics and Astronomy,
Rutgers University,}}
\centerline{\it Piscataway, NJ 08855-0849, USA}
\centerline{$^\flat${\it Theory Division, CERN, Geneva 23, CH-1211, Switzerland}}
\bigskip
\bigskip
\bigskip
\bigskip
\noindent
We compute all loop topological string amplitudes on orientifolds of local
Calabi-Yau manifolds, by
using geometric transitions involving SO/Sp Chern-Simons theory, localization on the moduli
space of holomorphic maps with involution, and the topological vertex.
In particular we count Klein bottles and projective planes with any number of handles 
in some Calabi-Yau orientifolds.

\vfill
\Date{May 2004}

\newsec{Introduction}

The large $N$ duality between open and closed topological strings, which
was first formulated for local conifold transitions \GV,
has been extended in various directions. It has been shown to be valid for more general
toric geometries,
leading to the definition of the topological vertex \AKMV, a cubic field theory
that computes the all genus
amplitudes of open and closed topological strings on any non-compact toric Calabi-Yau threefold. The large $N$ duality
has also been extended in \SV\ to a simple orientifold theory, namely the
orientifold of the conifold.

In this paper we propose a generalization of the large $N$
correspondence of \SV\ for more complicated orientifolds.
Namely, we find that the partition function of closed topological strings on the
orientifold (including unoriented
contributions and oriented contributions from the covering space) is equivalent in
the large $N$ limit to the
Chern-Simons partition function on the threefold after a geometric
transition. The ${\IP^1}$'s that were invariant under
the involution, becoming ${\IR\IP^2}$'s in the orientifold, give $SO(N)$ -
or $Sp(N)$ - Chern-Simons theory on the ${\bf S}^3$'s
resulting from the geometric transition. One has also to add
instanton contributions localized on the fixed
locus of a torus action on the deformed geometry.

This is a highly non-trivial proposal, as more complicated orientifolds involve
instanton contributions to the
Chern-Simons partition function. Moreover, for more general orientifolds, the
geometry of the covering space becomes
quite different from the one of the resulting orientifold. It is not obvious at
all that both the oriented and
unoriented contributions to the closed topological strings partition
function are encoded in the Chern-Simons setup.
But it turns out to be true in the examples we consider.

We also find that the
closed topological string amplitudes on the orientifolds of the type we describe below
can be also computed with the topological vertex introduced in \AKMV, by
using a prescription that takes into account the involution of the target.
We explicitly prove that this prescription is equivalent to the large $N$ Chern-Simons
dual. This prescription extends the general
formalism of the topological vertex to include the case of orientifolds.

To test our result we compute the unoriented contributions on the closed topological
strings side using the unoriented
localization techniques developed in \DFM.  This computation does not rely on large $N$ duality at all,
consequently providing an independent check of our proposal. In \SV\ it was found
that only unoriented maps with
one crosscap contribute to the partition function. However, in the general case, we find that
configurations with two crosscaps,
that is Klein bottles, do contribute as well.

To make the proposal more concrete we focus on a particular geometry in this paper. We consider a noncompact
Calabi-Yau threefold $X$ whose compact locus consists of two compact divisors each
isomorphic to a del Pezzo surface $dP_2$
and a rational $(-1,-1)$ curve that intersects both divisors transversely. The divisors
do not intersect each other. We
will equip $X$ with a freely acting antiholomorphic involution $I$ and consider an orientifold of the theory obtained
by gauging the discrete symmetry $\sigma I$, where $\sigma$ is a worldsheet antiholomorphic involution. The geometry is
described in more detail in section $3$.

The partition function of the closed topological ${\bf A}$-model with this geometry as
target space will sum both over
maps from orientable worldsheets to $X$ (with the K\"ahler parameters
identified by the involution set equal) as well as
over non-orientable worldsheets to the orientifolded geometry.

The orientifolded geometry allows a local geometric transition that will be described in detail in section $3$. This amounts
to contracting two ${\IP^1}$'s and an ${\IR\IP^2}$ and replace them by three $S^3$'s. We conjecture that the dual open string model
will consist of a system of Chern-Simons theories supported on the three spheres, with $U(N_1)$ and  $U(N_2)$ groups on the spheres
corresponding to the contracted $\IP^1$'s and $SO(N_3)$ - or $Sp(N_3)$ - group on the sphere corresponding to the contracted
$\IR\IP^2$. The new ingredient is that the Chern-Simons theories will be coupled by
cylindrical instantons.

The paper is organized as follows. Section $2$ summarizes general results for
{\bf A}-model topological strings on an
orientifold. Section $3$ describes in some detail the particular
geometry and the geometric transition on which we will focus in this paper.
In section $4$ we compute explicitly the Chern-Simons amplitude obtained after
the geometric transition. Section $5$ presents the unoriented localization
computation, and shows that it gives exactly the same contributions for the
one and two crosscaps instanton configurations. We then
propose our prescription based on the topological vertex in section $6$, proving
its equivalence to the Chern-Simons computation. Finally, in section
$7$, we point out some possible extensions of our results to more complicated
situations.

\newsec{{\bf A}-model Topological Strings on an Orientifold}

\subsec{Type IIA superstrings and topological strings on an orientifold}

It is a well-known fact that, when type IIA theory is compactified on a
Calabi-Yau manifold $X$, the resulting
four dimensional theory is ${\cal N}=2$ supergravity with $h^{1,1}(X)$ vector multiplets
$t_i$.
The ${\cal N}=2$ prepotential that governs the effective action of the vector multiplets,
$F_0(t_i)$, can be
computed by the genus zero free energy of the
{\bf A}-model topological strings with the Calabi-Yau as target space (see
\Hal\ for a review of topological strings and related issues). Higher genus
free energies $F_g(t_i)$ of the topological string theory also play a role in the
four dimensional supergravity theory, and compute higher curvature
couplings involving the graviphoton \refs{\BCOV, \AGNT}.

One way to break ${\cal N}=2$ supersymmetry down to ${\cal N}=1$ is
to consider an orientifold of the theory.
The orientifold is defined by combining an involution symmetry $I$ on
the Calabi-Yau $X$ with an orientation reversal
diffeomorphism $\sigma$ on the worldsheet $\Sigma$.
In the context of type IIA superstrings, the orientifold is only well defined if
the involution is anti-holomorphic.
Furthermore, the worldsheet diffeomorphism has to be orientation reversal
\refs{\AAHV,\DFM, \BH}. The resulting theory in four dimensions
has ${\cal N}=1$ supersymmetry, and $h_-^{1,1}(X)$ out of the $h^{1,1}(X)$ ${\cal N}=2$
vector multiplets
become ${\cal N}=1$ chiral multiplets in four dimensions,
where $h_-^{1,1}(X)$ is the number of
harmonic $(1,1)$ forms on $X$ which have $-1$ eigenvalue under $I$ (see \BH\ for
a description of the spectrum of massless modes in four dimensions).

These considerations hold in the context of {\bf A}-model topological strings as well:
{\bf A}-model topological strings possess a worldsheet orientation reversal
symmetry when accompanied with an
anti-holomorphic involution of the target space \AAHV.
It is thus possible to consider {\bf A}-model topological strings on an orientifold
defined as above. The twisted sector of the topological string amplitude on the
orientifold includes amplitudes for unoriented Riemann surfaces. Recall that
a closed, non-orientable
Riemann surface is characterized by its genus $g$ and by the
number of crosscaps $c$, which can be one or two (crosscaps can be
traded for handles when the number of crosscaps is higher than two).
For example, the surface with $g=0$ and $c=1$ is the real projective plane
$\IR \IP^2$, while the surface with $g=0$ and $c=2$ is the Klein bottle.
It was shown in \AAHV\ that the superpotential of the $h_-^{1,1}(X)$
chiral multiplets is given by the $\IR\IP^2$ amplitude of the topological
string theory. As far as we know, the topological amplitudes
involving more handles or crosscaps do not have an interpretation in the ${\cal N}=1$
supergravity theory.

Generally speaking, one could consider type IIA superstrings on a non-compact
orientifold, with D-branes and
orientifold planes \AAHV. In this paper we will only consider type IIA superstrings
without D-branes or orientifold planes
(albeit our approach could probably be generalized to these cases). This means
that the anti-holomorphic involution must have no fixed points. Moreover,
as the parent theory has no D-branes,
to compute the superpotential we only need to consider {\bf A}-model
closed topological strings.

\subsec{Structure of the topological string amplitudes}

Roughly speaking, the free energy of {\bf A}-model closed topological strings
counts the number of
holomorphic maps from the worldsheet to the target space, weighted by a
factor of $e^{-A}$ where $A$ is the area of the embedded curve.
In the context of orientifolds, the partition function of topological strings
sums over holomorphic maps in two different sectors:
the ``untwisted'' and the ``twisted'' sectors. The former consists of
usual holomorphic maps from orientable worldsheets to the covering space,
i.e. the non-compact Calabi-Yau threefold without the involution. The latter
consists of equivariant maps
$f: \Sigma \rightarrow X$ satisfying the equivariance
condition
\eqn\equiv{
f\circ \sigma = I \circ f,
}
where $I$ is the antiholomorphic involution acting on $X$, and
$\sigma: \Sigma \rightarrow \Sigma$ is the orientation
reversal diffeomorphism of the
Riemann surface which is needed in order to construct the orientifold
action. Notice that, if $\Sigma$ has genus zero, the action of
$\sigma$ is given by $z \rightarrow -1/{\bar z}$. The relevant maps in the
twisted sector are then the maps which
are compatible with the orientation reversal diffeomorphism on the worldsheet
and the anti-holomorphic involution on the target space, and
descend to holomorphic maps from non-orientable worldsheets to the orientifold.

The structure of the total free energy of the {\bf A}-model is then
\eqn\strucf{
\CF(X/I, g_s)= \CF(X/I, g_s)_{\rm or} + \CF(X/I, g_s)_{\rm unor},
}
where $g_s$ is the string coupling constant. In this equation,
$\CF(X/I, g_s)_{\rm or}$ is the contribution of the untwisted sector, and
$\CF( X/I, g_s)_{\rm unor}$ is the contribution of the twisted sector.
Moreover, we have \refs{\GViii,\SV}
\eqn\orcont{
  \CF(X/I, g_s)_{\rm or}={1\over 2}\CF(X, g_s)={1\over 2}
\sum_{d=1}^{\infty}\sum_{g=0}^{\infty}\sum_{Q} {1\over d} { n^g_Q \over (q^{d\over 2} - q^{-{d\over 2}})^{2-2g}}
e^{-dQ\cdot t}.
}
Here, $\CF(X, g_s)$ is the free energy of the covering $X$ of $X/I$,
after suitably identifying
the K\"ahler classes in the way prescribed by the involution $I$, and
we have written it in terms of Gopakumar-Vafa invariants $n^g_Q $ \GViii. The notation is
as follows: $t=(t_1, \cdots, t_n)$ denotes
the set of K\"ahler parameters of $X$ after identification through the involution,
$Q=(Q_1, \cdots, Q_n)$ is an $n$-uple of integer numbers that label integer
two-homology classes,
and $q=e^{g_s}$.

The unoriented contribution in \strucf\ comes from holomorphic maps from
closed non-orientable Riemann surfaces to the orientifold $X/I$.
The Euler characteristic of a closed Riemann surface of genus $g$ and $c$ crosscaps
is $\chi = -2g+2-c$ where $c$ is the number of crosscaps. We then have
\eqn\strucun{
\CF(X/I, g_s)_{\rm unor}= \CF(X/I, g_s)_{\rm unor}^{c=1}+ \CF(X/I, g_s)_{\rm unor}^{c=2},}
which corresponds to the contributions of one and two crosscaps. Following
the arguments in \GViii\ we expect the structure
\eqn\strucun{\eqalign{
\CF(X/I, g_s)_{\rm unor}^{c=1}=&
\pm \sum_{d \, \, {\rm odd}}\sum_{g=0}^{\infty}\sum_{Q }  n^{g,c=1}_Q
{1\over d}(q^{d\over 2} - q^{-{d\over 2}})^{2g-1}e^{-d Q\cdot t},\cr
\CF(X/I, g_s)_{\rm unor}^{c=2}=&
\sum_{d \, \, {\rm odd}} \sum_{g=0}^{\infty}\sum_{Q }
n^{g,c=2}_Q {1\over d}(q^{d\over 2} - q^{-{d\over 2}})^{2g}e^{-d Q\cdot t},\cr}
}
where $n^{g,c}_Q$ are integers. The $\pm$ sign in the $c=1$ free energy is due to
the following:
the target space anti-holomorphic involution does not fully specify the unoriented part
of the free energy
on the orientifold, since we have to make a choice for the sign of the crosscaps. Depending
on this choice, we will have
the two different signs for $c=1$. This corresponds to the choice of $SO$ or $Sp$ group in the gauge
theory dual. This remaining choice is also easily understood on the mirror symmetric
side \AAHV. For
the conifold, the
{\bf B}-model mirror symmetric description involves two orientifold 5-planes. The two
choices of signs for crosscap states correspond
on the mirror symmetric side to the two following choices for the charges of the
$O5$-planes: $+-$ and $-+$ \AAHV. A similar story holds for more complicated orientifolds. Notice as well that the sum
over multicoverings $d$ in \strucun\ is only over {\it odd} integers. In the case of
$c=1$ this follows from an elementary geometric argument, since there are no even
multicoverings (see \refs{\SV,\AAHV}). For $c=2$ there is no such a simple argument, but
our explicit computations both in Chern-Simons theory and in localization of unoriented
instantons indicate that only odd multicoverings contribute.

\newsec{Geometric Transitions} \seclab\geometry

\subsec{Orientifold of the resolved conifold and its geometric
transition}

In \SV\ it was proposed that in the large $N$ limit, closed topological strings on the
orientifold of the conifold are
dual to $SO(N)$/$Sp(N)$ Chern-Simons theory on ${\bf S}^3$, where the choice of gauge group is
related to the choice of sign for the crosscaps.
Since this is the starting point for our discussion, let us review in some
detail the results of \SV.

We start with a theory of topological
open strings on the deformed conifold defined by $z_1 z_4 - z_2 z_3=\mu$. The conifold
contains an ${\bf S}^3$, and if we wrap $2N$ branes on the
three-sphere, the spacetime description of the
open topological string theory is Chern-Simons theory on ${\bf S}^3$
with gauge group $U(2N)$ and at level $k$ (the level is related to the open string
coupling constant). We
now consider the following involution of the geometry
\eqn\stinv{
I:(z_1, z_2, z_3, z_4) \rightarrow (\bar z_4, -\bar z_3, -\bar z_2, \bar z_1)}
that leaves the
${\bf S}^3$ invariant. The string field
theory for the resulting open strings is now
Chern-Simons theory with gauge group $SO(N)$ or $Sp(N)$, depending on the choice of
orientifold action on the gauge group. The total free energy of the
Chern-Simons theory with gauge group $SO/Sp$ can be written as
\eqn\sopcs{
\CF=-\log S_{00}^{SO(N)/Sp(N)} = {1\over 2} \sum_{d=1}^{\infty} {1 \over d} {e^{-dt}\over (q^{d\over2}
-q^{-{d\over 2}})^2} \mp  \sum_{d \, {\rm odd}}
 {1 \over d} {e^{-dt/2}\over q^{d\over2} -q^{-{d\over 2}}},
}
where the $\mp$ sign corresponds to $SO/Sp$, respectively. In \sopcs, $q=e^{g_s}$, with
\eqn\stringcou{
g_s= { 2\pi i \over k + y},}
and $y$ is the dual Coxeter of the gauge group, which is $N-2$ for $SO(N)$ and
$N+1$ for $Sp(N)$.
The parameter $t$ in \sopcs\ is the
't Hooft parameter, given by
\eqn\thooft{
t=(N\mp 1) g_s,}
for $SO/Sp$, respectively.
\ifig\orienticon{Geometric transition for the orientifold of the conifold. The cross in the figure to the 
left represents an $\IR\IP^2$ obtained by quotienting a $\IP^1$ by the involution $I$, and the dashed line in 
the figure on the right represents an ${\bf S}^3$ with $SO/Sp$ gauge group. }
{\epsfxsize4.0in\epsfbox{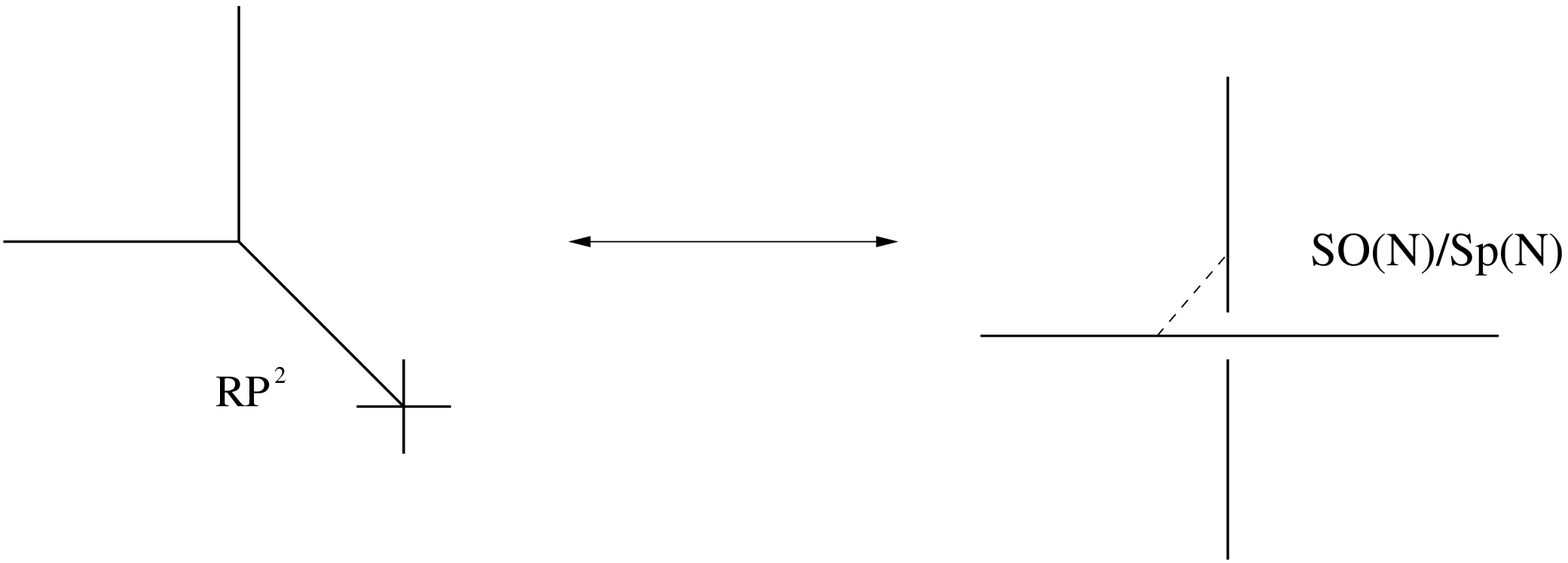}}
In the usual geometric transition of \GV, the dual to the deformed conifold is the
resolved conifold $Y={\cal O}(-1) \oplus {\cal O}(-1) \rightarrow {\IP}^1$. This Calabi-Yau
threefold
admits a toric description given by the following toric data:
\eqn\toricres{
\matrix{ &  X_1 & X_2 & X_3 & X_4  \cr
\IC^* & 1 & 1 & -1 & -1 }}
This means that $Y$ is defined as the space obtained from
\eqn\rescon{
|X_1|^2 + |X_2|^2 -|X_3|^2 - |X_4|^2 =t
}
after quotienting by the $U(1)$ action specified by the charges in \toricres.
The involution \stinv\ of the deformed conifold maps to the antiholomorphic
involution of $Y$ defined by:
\eqn\orres{
I: (X_1, X_2, X_3, X_4) \rightarrow ({\overline X}_2, -{\overline X}_1,
{\overline X}_4, -{\overline X}_3).
}
It is easy to see that $Y/I$ contains a single ${\IR\IP}^2$ obtained from the
quotient of the ${\IP}^1$ of $Y$ by $I$. We will represent the quotient of
the resolved conifold by that involution in terms of the toric diagram depicted 
in \orienticon.

The free energy of the $SO/Sp$ Chern-Simons theory gives the total free
energy of closed strings propagating on $Y/I$.
The first term in \sopcs\ gives the oriented contribution, while the
second term gives the unoriented contribution, and
they have the structure explained in \orcont\ and \strucun. Notice that
in the case of the unoriented contribution we have
\eqn\gvunor{
n_{Q=1/2}^{g=0,c=1}=\mp 1 }
depending on the choice of sign for the crosscaps, and all the remaining
Gopakumar-Vafa invariants vanish. In particular, the contribution
of Riemann surfaces with two crosscaps is zero. As we will see in this paper, in more general 
cases there are two-crosscaps contributions.
The above prediction of the large $N$ transition for the free energy was
checked in \AAHV\ against mirror symmetry, and in \DFM\ against
localization computations for unoriented Gromov-Witten theory.

\subsec{Our main example}

In this paper we want to generalize the open/closed string duality studied in
\SV\ to more general orientifolds. We will mainly focus on a noncompact
Calabi-Yau manifold $X$
described by the following toric data:
\eqn\toricXi{
\matrix{ &  X_1 & X_2 & X_3 & X_4 & X_5 & X_6 & X_7 & X_8 & X_9 & X_{10} \cr
\IC^* & -1 & 1 & 1 & -1 & 0 & 0 & 0 & 0 & 0 & 0 \cr
\IC^* & 1 & 0 & -1 & -1 & 0 & 1 & 0 & 0 & 0 & 0 \cr
\IC^* & 1 & -1 & 0 & -1 & 1 & 0 & 0 & 0 & 0 & 0 \cr
\IC^* & 0 & 0 & 0 & 1 & -1 & -1 & 1 & 0 & 0 & 0 \cr
\IC^* & 0 & 0 & 0 & 0 & 1 & 0 & -1 & -1 & 0 & 1 \cr
\IC^* & 0 & 0 & 0 & 0 & 0 & 1 & -1 & 0 & -1 & 1 \cr
\IC^* & 0 & 0 & 0 & 0 & 0 & 0 & -1 & 1 & 1 & -1.}}

\ifig\geom{Toric diagram of the noncompact Calabi-Yau threefold $X$.}{\epsfxsize2.0in\epsfbox{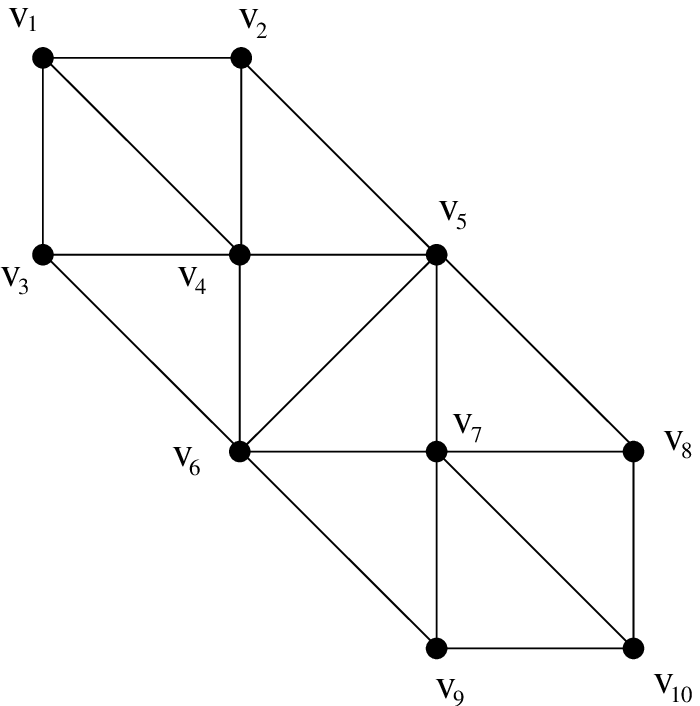}}

\noindent The compact locus consists of two divisors that are each isomorphic to a del Pezzo surface $dP_2$ and
a rational $(-1,-1)$ curve that intersects both divisors transversely. Note that the two compact divisors do
not intersect. We now consider a real torus action on $X$ given by:
\eqn\tact{\eqalign{
e^{i\phi}\cdot (X_1,X_2,\ldots,X_{10})\rightarrow
(e^{i\lambda_1\phi}X_1,e^{i\lambda_2\phi}X_2,\ldots,e^{i\lambda_{10}\phi}X_{10}).}}
The configuration of invariant curves is presented in fig. 3. We now define the antiholomorphic involution as follows:
\eqn\antinv{\eqalign{
I:(X_1,X_2,X_3,X_4,X_5,X_6,X_7,&X_8,X_9,X_{10})\rightarrow\cr
&(\bX_{10},\bX_8,\bX_9,\bX_7,-\bX_6,\bX_5,-\bX_4,\bX_2,\bX_3,\bX_1).
}}
The subtorus of \tact\ that is compatible with the involution is defined by the following constraints on the weigths
\eqn\constr{
\lambda_1+\lambda_{10}=0,~~\lambda_2+\lambda_8=0,~~\lambda_3+\lambda_9=0,~~\lambda_4+\lambda_7=0,~~
\lambda_5+\lambda_6=0.
}
Imposing these constraints does not enlarge the set of invariant curves.

It is often useful \AMV\ to consider a related Calabi-Yau threefold $\widetilde X$ obtained from $X$ by flopping
the two exceptional curves outside of the compact divisors. The ``commuting square'' of geometries (where the arrows
correspond either to flopping or to quotienting by the antiholomorphic involution) is presented in fig. 3.

\ifig\toric{The geometry on the closed topological strings side. The orientifolding action acts from left to right,
while flopping the ${\IP^1}$'s acts top-down. The ${\IR\IP^2}$ is represented by a cross at the end of the toric leg.}
{\epsfxsize4.5in\epsfbox{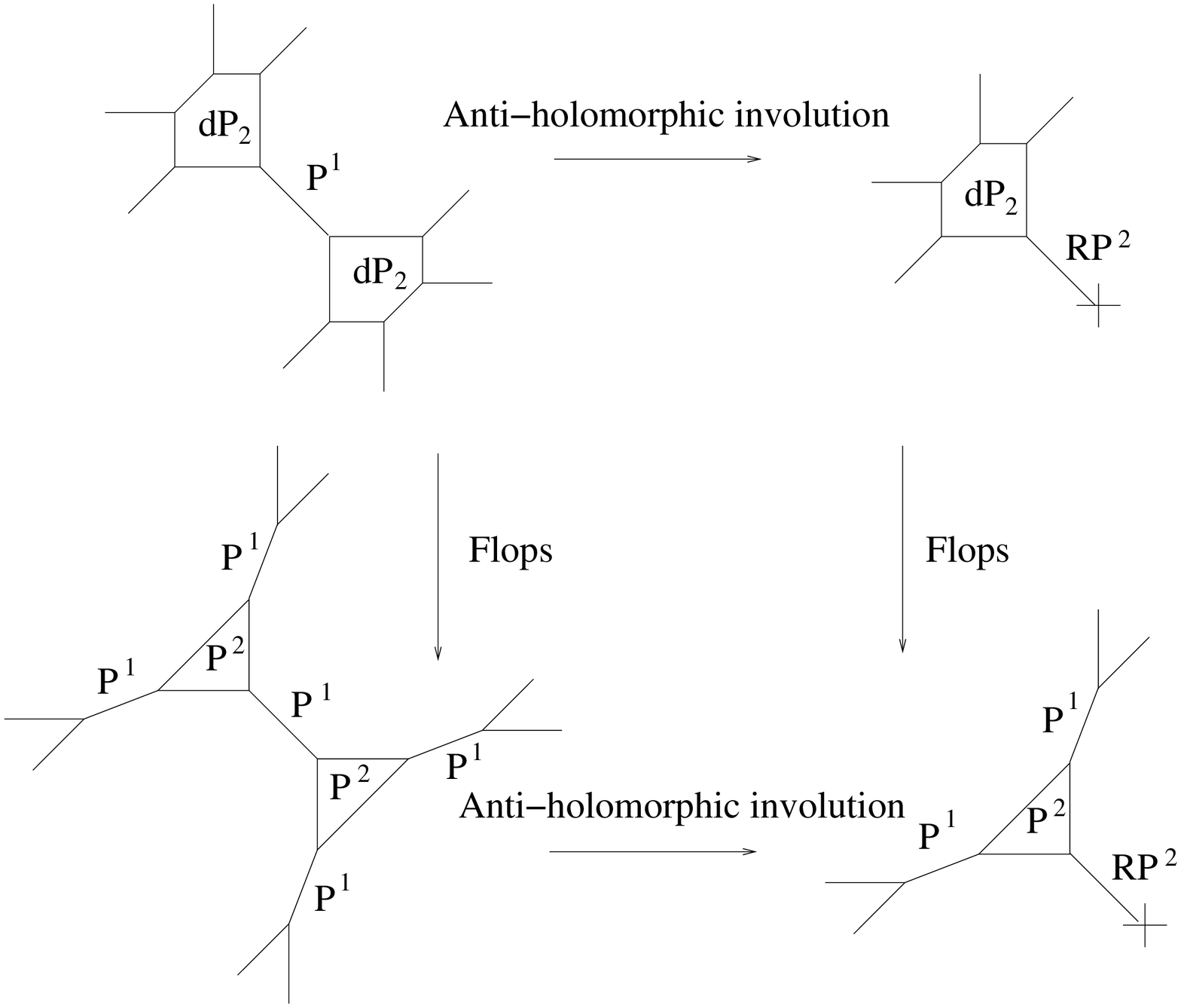}}

We can now follow the logic in \refs{\AV,\DFGi,\DFGii,\AMV} and consider a
geometric transition in which each of the resolved conifolds (or their orientifolds)
that exist locally in the geometry are replaced by
deformed conifolds (or their orientifolds). In the above example, this means that
we contract two $\IP^1$'s and an $\IR\IP^2$ and we replace them
with three spheres carrying $U(N)$ and $SO(N)/Sp(N)$ Chern-Simons theories, respectively.
The transition is represented
in \toric. In the next section we will see how to obtain the closed string amplitudes
in the orientifold from Chern-Simons theory

\ifig\transition{The geometric transition. The two ${\IP^1}$ and the ${\IR\IP^2}$ of
the left figure are shrunk to singular
points in the middle diagram, and then deformed into three $S^3$.}
{\epsfxsize5.5in\epsfbox{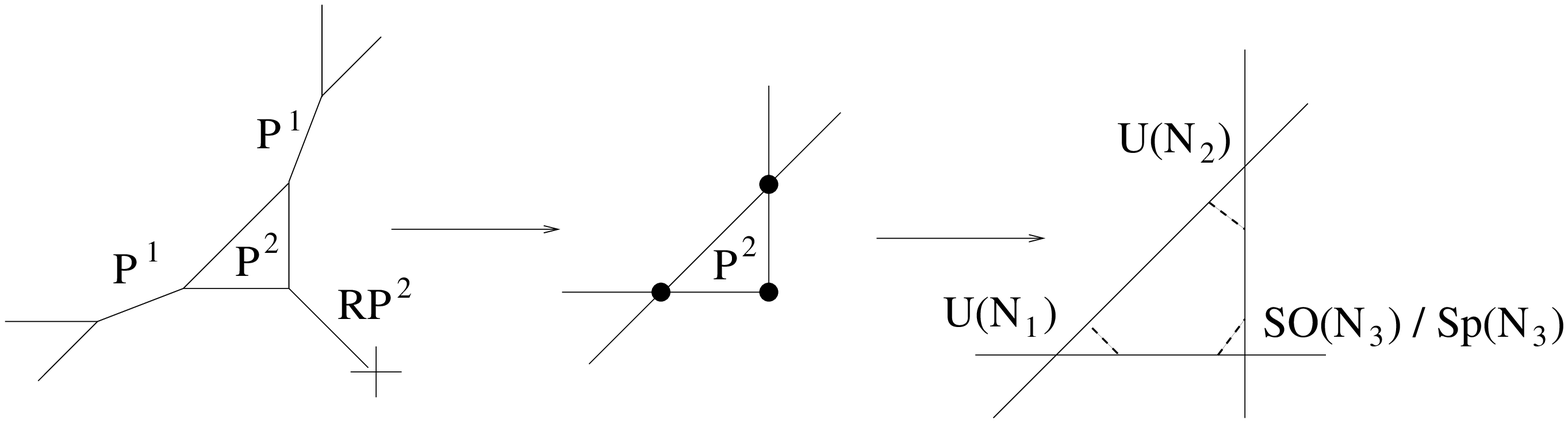}}

\newsec{Closed String Amplitudes from Chern-Simons Theory}

\subsec{Results from Chern-Simons theory with classical gauge groups}

As we will see in a moment,
in order to compute the free energies of topological strings on orientifolds via geometric
transitions we have to compute the Chern-Simons invariants of the
unknot and the Hopf link of linking number $+1$ in arbitrary representations
of $U(N)$, $SO(N)$ and $Sp(N)$.
In this section we will denote $q=e^{g_s}$ and
\eqn\lam{
\lambda=q^{N+a},}
where
\eqn\avalues{
a=\cases{0\,\,\,\,\,\,\,\,\,\,\,\,\,\,\,\,\,\,\,\,\,\,\,\,\,\,\, {\rm for
\,\,\,\,} U(N), \cr
-1\,\,\,\,\,\,\,\,\,\,\,\,\,\,\,\,\,\,\,\,\,\, {\rm for
\,\,\,\,} SO(N),\cr
1\,\,\,\,\,\,\,\,\,\,\,\,\,\,\,\,\,\,\,\,\,\,\,\,\,\,\, {\rm for
\,\,\,\,} Sp(N).}}
Notice that the 't Hooft parameter of the classical gauge groups
can be written as
\eqn\thooftgen{
t=(N+a)g_s
}
therefore $\lambda=e^t$.
For an arbitrary gauge group $G$ it is a well known result that the Chern-Simons invariant of the unknot in an arbitrary 
representation $R$ is given by the so-called {\it quantum dimension} of $R$ \Wiii:
\eqn\invunknot{
 \CW_R = {S_{0R} \over S_{00}} = \dim_q R,
}
where $S_{0R}$, $S_{00}$ are entries of the $S$ matrix of the WZW model with the corresponding gauge group and at level 
$k$ (recall that $k$ is related to the string coupling constant by \stringcou ). Using Weyl's formula one can write the 
quantum dimension as a product over positive roots $\alpha \in \Delta_+$:
\eqn\qdgeneral{
\dim_q R = \prod_{\alpha \in \Delta_+} {[(\Lambda_R + \rho, \alpha)] \over [(\rho, \alpha)]},
}
where $\Lambda_R$ is the highest weight of the representation $R$ and $\rho$ is the Weyl vector. We also defined the 
following $q$-number:
\eqn\qnumber{
[x]=q^{x/2} - q^{-x/2}.
}
The expression \qdgeneral\ can be written more explicitly for the different classical gauge groups.
Let $R$ be a representation corresponding to a Young tableau with row lengths $\{ \mu_i \}_{i=1,...,d(\mu)}$, with 
$\mu_1 \leq \mu_2 \leq ...$
and where $d(\mu)$ denotes the number of rows. Then the quantum dimension of a representation $R$ of $U(N)$ reads 
(see for example \MVi)
\eqn\qdUN{
\dim_q^{U(N)} R = \prod_{1 \leq i < j \leq d(\mu)} {[\mu_i-\mu_j+j-i] \over [j-i]} \prod_{i=1}^{d(\mu)} {\prod_{v=-i+1}^{\mu_i-1}
[v]_{\lambda} \over \prod_{w=1}^{\mu_i} [w-i+d(\mu)]},
}
where we defined
\eqn\qnumberii{
[x]_{\lambda} = \lambda^{1/2} q^{x/2} -\lambda^{-1/2} q^{-x/2},
}
and $\lambda=q^{N}$ for $U(N)$ representations.

We can also find explicit expressions for the quantum dimensions of $SO(N)$ and $Sp(N)$ representations
\eqn\qdSON{\eqalign{
\dim_q^{SO(N)} R =& \prod_{1 \leq i < j \leq d(\mu)} {[\mu_i-\mu_j+j-i] [\mu_i + \mu_j+1-i-j]_{\lambda} 
\over [j-i][1-i-j]_{\lambda}}\cr
&\times \prod_{i=1}^{d(\mu)}{[\mu_i-i]_{\lambda}^{SO(N)} \over [-i]_{\lambda}^{SO(N)}} \prod_{v=1}^{\mu_i}
{ [\mu_i+1-i-v-d(\mu)]_{\lambda}
\over  [v-i+d(\mu)]},\cr
\dim_q^{Sp(N)} R =& \prod_{1 \leq i < j \leq d(\mu)} {[\mu_i-\mu_j+j-i] [\mu_i + \mu_j+1-i-j]_{\lambda} \over 
[j-i][1-i-j]_{\lambda}} \cr
&\times \prod_{i=1}^{d(\mu)}{[1-i]_{\lambda}^{Sp(N)} [2\mu_i-2i+1]_{\lambda} \over [1-i+\mu_i]_{\lambda}^{Sp(N)} 
[1-2i]_{\lambda}}\prod_{v=1}^{\mu_i}{ [\mu_i+1-i-v-d(\mu)]_{\lambda} \over  [v-i+d(\mu)]},
}}
where we defined
\eqn\qnumberiii{\eqalign{
[x]_{\lambda}^{SO(N)} = \lambda^{1/4} q^{{1 \over 4} (2x+1)} -\lambda^{-1/4} q^{-{1\over 4}(2x+1)},\cr
[x]_{\lambda}^{Sp(N)} = \lambda^{1/4} q^{{1 \over 4} (2x-1)} -\lambda^{-1/4} q^{-{1\over 4}(2x-1)},
}}
with $\lambda=q^{N+a}$ which leads to $\lambda=q^{N-1}$ for $SO(N)$ and $\lambda=q^{N+1}$ for $Sp(N)$. Using \qdSON\ 
one can show that
\eqn\SOSpreln{
\dim_q^{Sp(N)} R = (-1)^{\ell(R)} \dim_q^{SO(-N)} R^T,
}
where $R^T$ is the transposed or conjugate 
representation, related to $R$ by exchanging rows with columns, $SO(-N)$ is meant in the sense of analytic
continuation,
and $\ell(R)$ is the number of boxes of the Young tableau.
This relation is part of the ``$SO(N)=Sp(-N)$'' equivalence \CVI. A relation similar to \SOSpreln\ holds
for usual dimensions \Ra.

Using \qdUN\ and \qdSON\ one can also infer the following formula for quantum dimensions of representations of $SO(N)$ 
and $Sp(N)$ in terms of quantum dimensions of representations of $U(N)$:
\eqn\qdSONUN{
\dim_q^{SO(N)/Sp(N)} R = \sum_{Q=Q^t}(-1)^{1/2(\ell(Q)\mp r(Q))} \dim_q^{U(N)} (R/Q),
}
where the skew quantum dimension is defined by
\eqn\skewq{
\dim_q^{U(N)} (R/Q)=\sum_{R'} N_{R' Q}^R  \dim_q^{U(N)} R'
}
and $N_{R_1 R_2}^R$ are the usual Littlewood-Richardson coefficients that appear in the tensor product of $U(N)$ representations:
$R_1 \otimes R_2 =\sum_R N_{R_1 R_2}^R R$. In \qdSONUN\ the sum is over self-conjugate representations, {\it i.e.} representations
that are equal to their transpose, and starts with the trivial representation: $\{ \cdot, \tableau{1}, \tableau{2 1},
\tableau{2 2}, \tableau{3 1 1}, ... \}$. 
$r(Q)$ denotes the rank of $Q$, which is defined as the number of boxes in the leading diagonal of the Young tableau \Li. The
$-$ sign is for $SO(N)$ representations while the $+$ sign is for $Sp(N)$ representations.

As we will see in the following sections, the relations between quantum dimensions of representations of
$SO(N)$ and $Sp(N)$ \SOSpreln\ and \qdSONUN\ are responsible for the fact that partition functions of $SO(N)$
and $Sp(N)$ differ only by an overall sign in front of the unoriented contributions with an odd number of crosscaps,
which lead to the interpretation that they correspond to different choices of sign for the crosscap states. 
Basically, the first term in the sum of the right hand side of \qdSONUN\
is responsible for oriented contributions to the partition functions, so they are the same for $SO(N)$, $Sp(N)$ and
$U(N)$ gauge groups. The other terms in the sum are responsible for unoriented contributions
to the partition function, and the difference of sign in the exponent of the $(-1)$ factor leads to a relative minus
sign between unoriented contributions with an odd number of crosscaps of the $SO(N)$ and $Sp(N)$ partition functions.

Another important ingredient we will need is the framing of knots and links \Wiii. Given a knot invariant
in representation $R$, we can change its framing
by $p$ units (where $p$ is an integer) if we multiply it by
\eqn\framing{
(-1)^{\ell(R) p}q^{p C_R/2}
}
where $C_R$ is the quadratic Casimir of the representation $R$. The quadratic Casimirs have the following expressions 
for the different classical gauge groups:
\eqn\crvalues{
C_R=\kappa_R + (N+a) \ell(R),
}
where $a$ is given by \avalues, and
\eqn\kappaquantity{
\kappa_R = \sum_i \mu_i (\mu_i - 2 i+1).
}
The framing factor can then be written as
\eqn\framingl{
(-1)^{\ell(R) p} \lambda^{p \ell(R)/2}q^{p \kappa_R/2}.
}
The sign in \framing\ is not standard in the context of Chern-Simons theory, but as shown in
\refs{\AKV,\MVi}, it is crucial in the context of topological string theory in order to guarantee integrality
properties in the resulting amplitudes. To incorporate a change of framing in a link, we just change the framings
of each of its components according to the rule \framing\ as well.

In our computations we will also need the invariants of Hopf links with linking number $+1$. For arbitrary gauge group $\CG$,
the invariant of the Hopf link with linking number $+1$ is given by the normalized 
inverse $S$ matrix \Wiii, and it can be written in terms of quantum dimensions as (see for example \guada)
\eqn\hopflinkgeneral{
\CW_{R_1 R_2} = {S_{R_1 R_2}^{-1} \over S_{00}} = \sum_{R \in R_1 \otimes R_2}  q^{{1\over2}(C_R-C_{R_1}-C_{R_2})} \dim_q R,
}
where the sum is over all representations $R$ occurring in the decomposition of the tensor product of $R_1$ and $R_2$.
In the $U(N)$ case, we can replace the Casimir operators $C_{R_i}$ appearing in \hopflinkgeneral\ by $\kappa_{R_i}$, since
$\ell(R)=\ell (R_1)+\ell (R_2)$ in the decomposition of a tensor product of irreducible representations of $U(N)$.
However this relation between the number of boxes of Young tableaux does not hold in the $SO(N)$ and $Sp(N)$ cases. We thus find
\eqn\hopflink{\eqalign{
\CW_{R_1R_2}^{U(N)} &=  \sum_{R} N_{R_1 R_2}^R  q^{{1\over2}(\kappa_R-\kappa_{R_1}-\kappa_{R_2})} \dim_q^{U(N)} R,\cr
\CW_{R_1R_2}^{SO(N)/Sp(N)} &=  \sum_{R } M_{R_1 R_2}^R
\lambda^{{1\over2}(\ell(R)-\ell(R_1)-\ell(R_2))} q^{{1\over2}(\kappa_R-\kappa_{R_1}-\kappa_{R_2})} \dim_q^{SO(N)/Sp(N)} R,
}}
where we have denoted by $M_{R_1 R_2}^R$ the tensor product coefficients for irreducible representations of
$SO(N)$ and $Sp(N)$, which turn out to be the same for $SO$ and $Sp$.

To compute \hopflink\ we need the values of $M_{R_1 R_2}^R$, in other words, we have to
decompose any tensor product of $SO(N)$ or $Sp(N)$ representations
into a sum of irreducible representations. This can be done with a technique first developped by Littlewood in \Li.
Let us first consider $SO(N)$ representations. Let $[R]$ be
the character of the representations $R$, as a function of the eigenvalues of an $SO(N)$ matrix,
and let $\{R\}$ be the Schur function of these eigenvalues labeled by the same representation.
One can prove the following formulae \Li:
\eqn\characSON{\eqalign{
[R]=&\{R\} + \sum_{R_1 \in \{\delta\}} (-1)^{\ell(R_1)/2} N_{R_1R_2}^R \{ R_2 \},\cr
\{R\}=&[R]+\sum_{R_1 \in \{\gamma\}}  N_{R_1R_2}^R [R_2],
}}
where $\{\delta\}$ and $\{\gamma\}$ are subsets of Young tableaux that we describe in Appendix A.
By using these relations one can express each character $[R][R']$ in the product as a sum of Schur functions, then multiply these
with the usual Littlewood-Richardson coefficients, and finally rexpress the Schur functions in terms of a sum of characters by the
second equation of
\characSON. For example,
\eqn\examptensor{\eqalign{
[\tableau{2}][\tableau{1}]=&(-1+\{\tableau{2}\})(\{\tableau{1}\})\cr
=&\{\tableau{3}\}+\{\tableau{2 1}\}-\{\tableau{1}\}\cr
=&[\tableau{3}]+[\tableau{1}]+[\tableau{2 1}]+[\tableau{1}]-[\tableau{1}]=[\tableau{1}]+[\tableau{3}]+[\tableau{2 1}],
}}
where the Young tableaux are associated to irreducible representations of $SO(N)$.
To compute the decompositions for $Sp(N)$ representations, one only has to replace the
subsets $\{\delta\}$ and $\{\gamma\}$ respectively by the subsets $\{ \beta \}$ and $\{ \alpha \}$, which are also explained 
in Appendix A \Ra. Using this technique one can decompose any tensor products of $SO(N)$ and $Sp(N)$ representations into a 
sum of irreducible representations, which is needed in the computation of expectation values of Hopf links using \hopflink. 
One finds that the decomposition of tensor products is always the same for $SO(N)$ and $Sp(N)$ representations, justifying 
our claim above.

The procedure we have described turns out to be rather involved, and fortunately there is a more direct way of computing 
$M_{R_1 R_2}^R$ through the following formula \refs{\king,\FH}:
\eqn\kingfor{
M_{R_1 R_2}^R = \sum_{Q,T,U} N_{QT}^{R_1} N_{QU}^{R_2} N_{TU}^R,
}
which expresses these coefficients in terms of usual Littlewood-Richardson coefficients. This formula allows to easily
compute the invariants of Hopf links for $SO/Sp$ gauge groups for any pair of representations.

As shown in \refs{\AMV,\AKMV}, the Hopf link invariant $\CW_{R_1R_2}^{U(N)}$ plays a crucial role in the computation of
oriented string amplitudes. It is a Laurent polynomial in $\lambda^{1\over 2}$ whose highest power is $\lambda^{(\ell(R_1) 
+ \ell(R_2))/2}$:
\eqn\leadinghopf{
\CW_{R_1 R_2}^{U(N)}=\lambda^{(\ell(R_1) + \ell(R_2))/2} W_{R_1 R_2}(q)+ \cdots,}
where the dots refer to terms with lower powers of $\lambda$.
The leading part of $\CW_{R_1 R_2}^{U(N)}$, which we have denoted by $W_{R_1 R_2}$, can be
computed in terms of Schur polynomials in an infinite number of
variables (see for example \refs{\AKMV,\ORV,\EK} for more details):
\eqn\hopfschur{
W_{R_1 R_2}(q)= s_{R_2}(x_i=q^{-i+{1\over 2}})s_{R_1}(x_i=q^{\mu_i^{R_2}-i+{1\over 2}}),
}
where $\{ \mu_i^{R_2} \}_{i=1, \cdots, d(\mu^{R_2})}$ is the partition corresponding to $R_2$. 
We will also denote $W_R=W_{R \cdot}=s_R (x_i=q^{-i+{1\over 2}})$.
By looking at the formula in \hopflink\ for $\CW_{R_1R_2}^{SO(N)/Sp(N)}$, one can see that it is a
Laurent polynomial in $\lambda^{1\over 2}$, whose highest power is also $\lambda^{(\ell(R_1) + \ell(R_2))/2}$, and
which has the same leading coefficient $W_{R_1 R_2}(q)$.

\subsec{Computation of open string amplitudes}

\ifig\CSsetup{The deformed geometry. $M_i$ $i=1,2,3$ are the three spheres and
$r_{c_i}$ are the K\"ahler parameters of the cylindrical instantons.
The gauge groups of the Chern-Simons theories on the spheres and the framings of the unknots are also indicated.}
{\epsfxsize3.5in\epsfbox{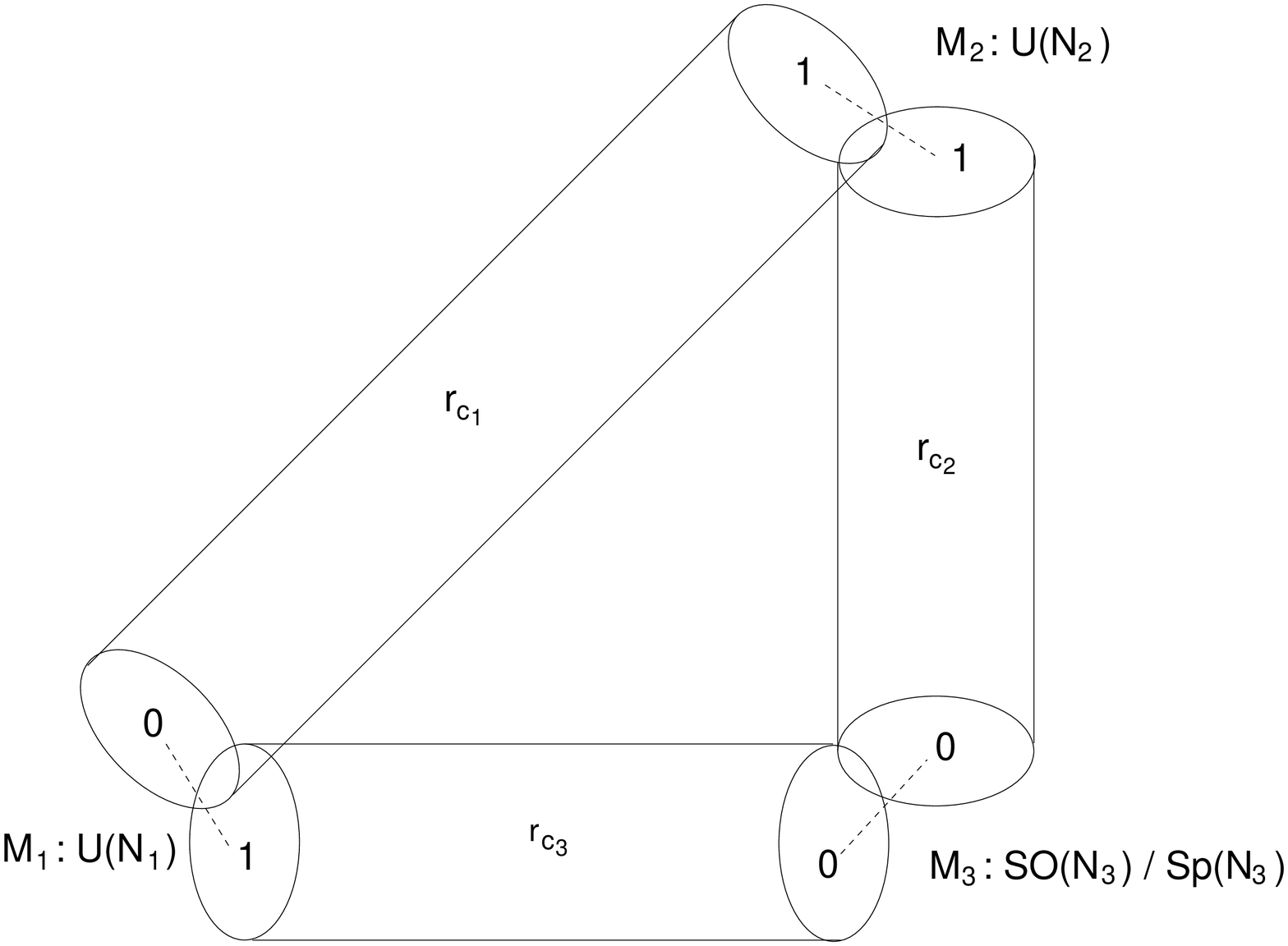}}
We will now follow the results in \refs{\AMV,\DFGi,\DFGii} to compute the topological open string amplitudes
in the geometry described in section \geometry,
which is shown in \CSsetup. There are $N_i$ D-branes wrapped around the three ${\bf S}^3$'s $M_i$, $i=1,2,3$.
This geometry is similar to the one considered for example in \AMV; the main difference being that one
of the spheres in our geometry, more precisely $M_3$, is left invariant by the anti-holomorphic involution,
thus leading to a $SO(N)$ or $Sp(N)$ Chern-Simons theory.

For open strings with both ends on the same ${\bf S}^3$, the dynamics is described by a Chern-Simons theory as usual.
For $M_1$ and $M_2$, the Chern-Simons theories respectively have gauge groups $U(N_1)$ and $U(N_2)$, while for $M_3$
it has gauge group $SO(N_3)$ or $Sp(N_3)$. However, there are also cylindrical open string instantons coupling the
Chern-Simons theories on different spheres \Wii. Schematically, the path integral becomes
\eqn\PI{
Z=\int \prod_{i=1}^3 {\cal D} A_i \, e^{\sum_{i=1}^3 S_i^{CS}(A_i) + S_{inst}},}
where $S_i^{CS} (A_i)$, $i=1,2,3$ are the Chern-Simons actions for the three ${\bf S}^3$'s. The instanton sector,
$S_{inst}$, can be computed by using localization (as in \refs{\DFGi,\DFGii}) or by using the techniques of \AMV.
We will follow here the procedure of \AMV. As explained there,
the bifundamental strings stretching between two three sphere ${\bf S}^3$'s give a massive complex scalar field, with
mass proportional to the complexified K\"ahler parameter $r_c$ corresponding to the ``distance'' between the two spheres.
After integrating out this scalar field one finds an operator which corresponds to a primitive annulus of size $r_c$ 
together with its multicovers. The boundaries of the annulus are on the two three spheres between which the 
bifundamental strings are stretched. These cylindrical instantons and the geometry are shown in \CSsetup. 
Inserting one operator for each cylindrical instanton we find
\eqn\Finst{
e^{S_{inst}}= \CO(U_3,U_1) \CO(V_1, V_2) \CO(U_2,V_3),
}
where we have defined the holonomy variables
\eqn\holonomy{
U_i=P \exp \oint_{\Xi_i} A_i,~~~V_i = P \exp \oint_{\Gamma_i} A_i,~~~i=1,2,3,
}
and the $\Xi_i,\Gamma_i$, $i=1,2,3$ are the boundary components of the cylindrical instantons, which are unknots in the
corresponding three-spheres. The operators in \Finst\ are given by
\eqn\operatorR{
\CO(A,B, r_c)= \sum_{R}\Tr_R A e^{-\ell(R) r_c} \Tr_R B,
}
where the sum is over all representations, including the trivial one.

The careful reader may note that the operator \operatorR\ is only equivalent to the
usual operator \refs{\OV,\AMV}
\eqn\operatorOV{
\exp \sum_{n=1}^{\infty} {e^{-n r_c} \over n} \Tr A^n  \Tr B^{n}
}
in the $U(N)$ case. In the more general case where the gauge group is $SO(N)$ or $Sp(N)$,
the two operators are not equivalent. It turns out that \operatorR\ is the good operator
to use; it would be interesting to investigate further why this is so.

We can now write the total free energy $\CF=-\log Z$ (with $Z$ given in \PI) as
\eqn\freeP{
\CF=\sum_{i=1}^3 \CF(M_i) + {\cal F}_{inst},
}
where $\CF(M_i)$ are the free energies of the Chern-Simons theories in the spheres $M_i$,
$i=1,2,3$, and ${\cal F}_{inst}$ is:
\eqn\Finstii{
\CF_{inst}= - \ln \biggl\{
\sum_{R_1,R_2,R_3}  e^{-\sum_{i=1}^3 \ell(R_i) r_{ci}} W_{R_3 R_1} (\CL_1) W_{R_1 R_2} (\CL_2) W_{R_2 R_3} (\CL_3)
\biggr\},
}
where $\CL_i$ is the link formed by the knots $(\Xi_i,\Gamma_i)$ and
\eqn\Wdef{\eqalign{
W_{R_3R_1} (\CL_1)&={\langle R_{3} | V_{M_{1}} | R_1 \rangle \over Z_{M_{1}}} ,\cr
W_{R_1R_2} (\CL_2)&={\langle R_{1} | V_{M_{2}} | R_2 \rangle \over Z_{M_{2}}},\cr
W_{R_2R_3} (\CL_3)&={\langle R_{2} | V_{M_{3}} | R_3 \rangle \over Z_{M_{3}}} .
}}
It was shown in \AMV\ (using our notation as in \CSsetup) that
\eqn\Vmatrix{
V_{M_1} = TS^{-1},~~~V_{M_2} = S T^{-1} S,~~~ V_{M_3}=S^{-1},
}
which means that the three links $\CL_i$, $i=1,2,3$ are Hopf links with linking number $+1$ and that the framings 
are as follows:
$(\Gamma_1, \Xi_3, \Gamma_3)$ are canonically framed, i.e. with framings $(0,0,0)$, while $(\Xi_1,\Xi_2,\Gamma_2)$
have framings $(1,1,1)$, as shown in \CSsetup. We can thus write
\eqn\Ws{\eqalign{
W_{R_3 R_1} (\CL_1)&=  (-1)^{\ell(R_3)} q^{\kappa_{R_3} \over 2} {S_{R_3 R_1}^{-1} \over S_{00}}=(-1)^{\ell(R_3)} 
q^{\kappa_{R_3} \over 2} \CW_{R_3R_1},\cr
W_{R_1 R_2} (\CL_2)&=  (-1)^{\ell(R_1)+ \ell(R_2)} q^{{1\over 2}(\kappa_{R_1}+\kappa_{R_2})} {S_{R_1 R_2}^{-1} 
\over S_{00}} =
(-1)^{\ell(R_1)+ \ell(R_2)} q^{{1\over 2}(\kappa_{R_1}+\kappa_{R_2})} \CW_{R_1 R_2}\cr
W_{R_2 R_3} (\CL_3)&= {S_{R_2 R_3}^{-1} \over S_{00}} =\CW_{R_1 R_3},
}}
where the $\lambda$ dependent pieces of \framingl\ have been absorbed in a redefinition of $r_{ci}$. Therefore 
\Finstii\ becomes

\eqn\Finstiii{\eqalign{
\CF_{inst}&= - \ln \biggl\{
 1+ \sum_{R_1,R_2,R_3} (-1)^{\sum_{i=1}^3 l_i} e^{-\sum_{i=1}^3 \ell(R_i) r_{ci}} q^{{1 \over 2}(\kappa_{R_1}+\kappa_{R_2}
+\kappa_{R_3})}\cr
&\times \CW_{R_3R_1}(\CL_1) \CW_{R_1R_2} (\CL_2) \CW_{R_2R_3} (\CL_3) \biggr\},
}}
where we singled out the term coming from $R_1,R_2,R_3=\cdot$, i.e. the three representations being the trivial representation.

\subsec{Duality map and closed string amplitudes}

Let us first recall the variables we have defined so far. We first defined the Chern-Simons variables $q=e^{g_s}$
and $\lambda_i=q^{N_i+a_i}$, with $g_s = {2 \pi i \over k_i+y}$ being the same for the three theories. We denote the 
three K\"ahler parameters of the cylindrical instantons by $r_{ci}$, $i=1,2,3$ and the three 't Hooft 
parameters of the different gauge groups by $t_i$. We now want to relate the open string parameters $t_i$ and $r_{ci}$ 
to the following closed string parameters: $t$, which is the K\"ahler parameter of ${\IP}^2$, and $s_i$, $i=1,2,3$, 
which are the K\"ahler parameters of the two ${\IP}^1$'s and the ${\IR\IP}^2$. The duality map reads
\eqn\duality{\eqalign{
&t = r_{c1} - {t_1+t_2 \over 2} =  r_{c2} - {t_2 + t_3 \over 2}= r_{c3} - {t_1 + t_3 \over 2},\cr
&t_1 = s_1,~~~t_2 = s_2,~~~t_3 = s_3 .
}}
Let now $q_i$ be $q_i=e^{-s_i}=e^{-t_i}$, $i=1,2,3$, and let $\ell$ be $\ell(R_1)+\ell(R_2)+\ell(R_3)$. We can rewrite the
open string partition function \Finstiii\ using \duality:
\eqn\Finstiv{\eqalign{
\CF_{inst}=& -\ln \biggl\{  1+ \sum_{\ell} (-1)^{\ell} e^{-\ell t} q^{{1 \over 2}(\kappa_{R_1}+\kappa_{R_2}+\kappa_{R_3})} 
q_1^{\ell(R_1)+
\ell(R_3) \over 2} q_2^{\ell (R_1) +\ell(R_2) \over 2} q_3^{\ell(R_2)+\ell(R_3) \over 2} \cr
&\times \CW_{R_3 R_1}^{U(N)} \CW_{R_1 R_2}^{U(N)} \CW_{R_2 R_3}^{SO(N)/Sp(N)}  \biggr\},
}}
where the Hopf link invariants in the last line are evaluated at $\lambda=q_i^{-1}$, $i=1,2,3$, respectively.
Notice that the leading power of $\lambda$ in $\CW_{R_1 R_2}^{U(N)}$ and in $\CW_{R_1 R_2}^{SO(N)/Sp(N)}$ is in both
cases $\lambda^{(\ell(R_1) + \ell(R_2))/2}$, therefore the above expression for $\CF_{inst}$ gives a power series in $q_i$ with
positive integer coefficients, as it should.
We can now expand the logarithm to find
\eqn\Finstv{
\CF_{inst}=\sum_{\ell=1}^{\infty} Z_{\ell}^{(c)} e^{-\ell t},
}
where the connected coefficient $Z_{\ell}^{(c)}$ are given by
\eqn\connected{
Z_{\ell}^{(c)}=\sum_{1 \leq d \leq \ell} {(-1)^{d+1} \over d} \sum_{m_1+m_2+...+m_d=\ell} Z_{m_1} Z_{m_2} \cdots Z_{m_d}.
}
These coefficients give the instanton partition function order by order in the K\"ahler parameter $e^{-t}$. Using the
formulae given above for Hopf link invariants with classical gauge groups, we can explicitly compute
the coefficients $Z_{\ell}^{(c)}$.
The contributions independent of the K\"ahler parameter $t$ are given by the sum of Chern-Simons free energies on ${\bf S}^3$
$\sum_{i=1}^3 \CF_{CS}(M_i)$,
which have already been computed in \refs{\SV, \GV}.

\subsec{The oriented contribution}
\ifig\topologicalfig{The geometry in the topological vertex formalism. In brackets next to the representations are the framings
in the corresponding propagator.}
{\epsfxsize4in\epsfbox{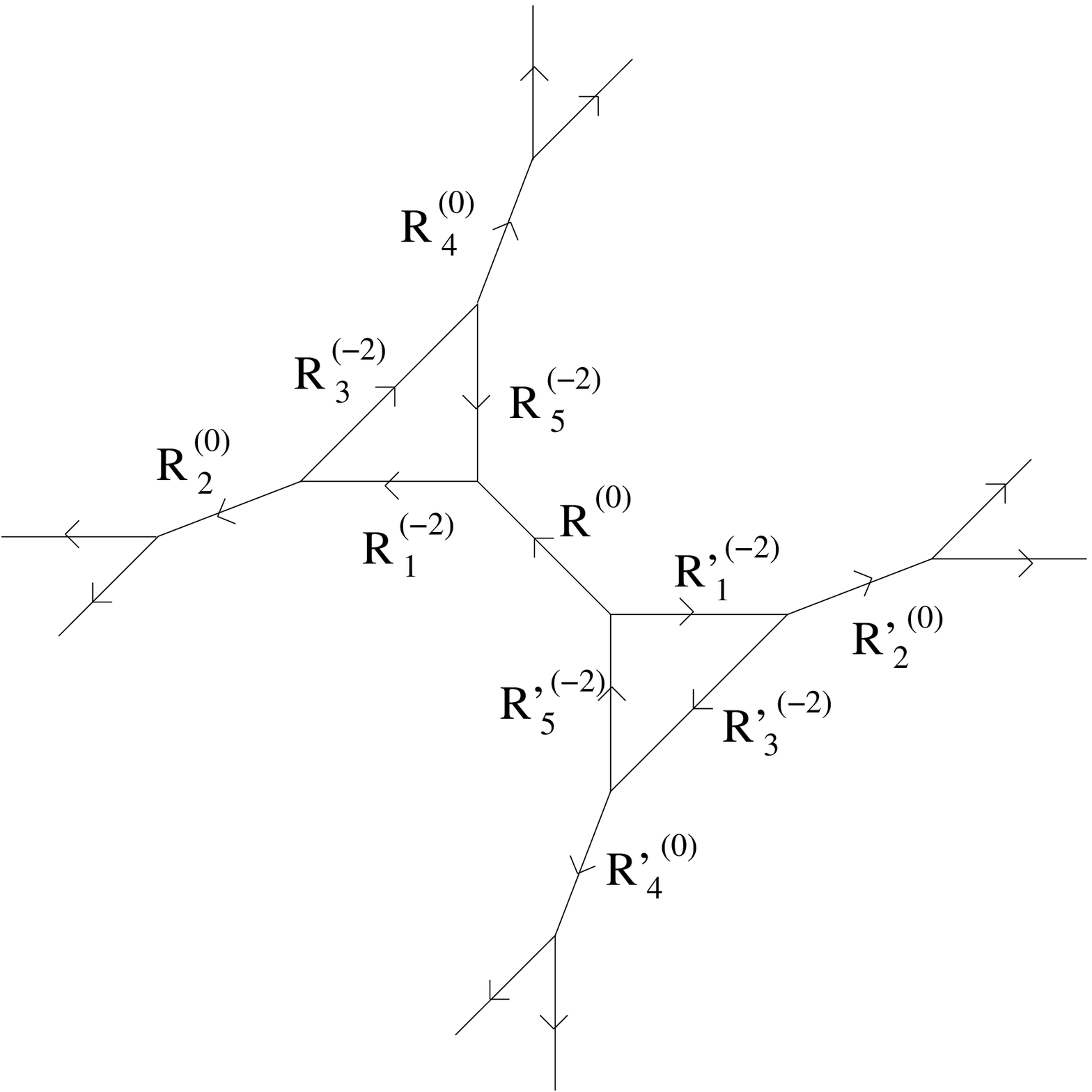}}
As we explained in section 2, $\CF_{inst}$ contains contributions due to oriented and to unoriented instantons.
In order to compute the closed, unoriented string amplitudes we have to subtract the oriented part, which we must compute
independently. The covering space $X$ is the Calabi-Yau manifold depicted in \topologicalfig. The oriented amplitude 
can be computed using the topological vertex formalism \AKMV. Using the formulas and gluing rules explained in \AKMV\ we find
\eqn\Ztopological{
Z(X)=\sum_{R} \CO_{R} (t, q_1, q_2) \CO_{R^T} (t_, q_1, q_2) (-1)^{l(R)} q_3^{l(R)},
}
where $q_i=e^{-s_{i}}$ and $t$ is the K\"ahler parameter of the ${\IP}^2$, $s_{1}$ and $s_{2}$ are the K\"ahler parameters of the
two ${\IP}^1$'s attached to the ${\IP}^2$, and $s_{3}$ is the K\"ahler parameter of the ${\IP}^1$ between the two ${\IP}^2$'s.
Notice that we have identified the K\"ahler parameters in the way prescribed by the involution.
In \Ztopological\ we introduced the operator
\eqn\Optopological{\eqalign{
\CO_{R} (t,q_1,q_2) =& \sum_{R_i} C_{R R_5 R_1^T} C_{R_1 R_3^T R_2^T} C_{R_3 R_5^T R_4^T} C_{R_2 \cdot \cdot} C_{R_4 \cdot \cdot}\cr
&\times (-1)^{\sum_{i} \ell(R_i)} q^{\kappa(R_1)+\kappa(R_3)+\kappa(R_5)}
e^{-(\ell(R_1)+\ell(R_3)+\ell(R_5))t} q_1^{\ell(R_2)} q_2^{\ell(R_4)},
}}
where $C_{R_i R_j R_k}$ is the topological vertex amplitude, which can be expressed in terms of the quantities \hopfschur:
\eqn\topvertex{
C_{R_1 R_2 R_3}=\sum_{Q_1,Q_3, Q} N_{Q Q_1}^{~~R_1} N_{Q
Q_3}^{~~R_3^T}
q^{\kappa_{R_2}/2+\kappa_{R_3}/2}
{W_{R_2^T Q_1}W_{R_2 Q_3}\over W_{R_2}}.
}
Using \Ztopological\ we can express again the free energy as a sum over connected coefficients
\eqn\Ftopological{
\CF (X)=-\log Z(X) = 
\sum_{\ell,\ell_1, \ell_2, \ell_3} Z_{\ell,\ell_1,\ell_2,\ell_3}^{(c)} q_1^{\ell_1} q_2^{\ell_2} q_3^{\ell_3} e^{-\ell t}.
}
The free energy computed in \Finstv\ should equal, according to \strucf,
\eqn\frees{
\CF_{inst}= {1\over 2}\CF(X) + \CF(X/I,g_s)_{unor},
}
where $\CF(X)$ is given in \Ftopological. This determines the unoriented part, which should have
the structure given in \strucun. We will
encode the resulting oriented and unoriented Gopakumar-Vafa invariants in the
following generating functionals
\eqn\functionalsi{\eqalign{
\CF_d^{g,0}&={1 \over 2} \sum_{d_1,d_2,d_3} n^{g,0}_{d,d_1,d_2,d_3} q_1^{d_1} q_2^{d_2} q_3^{d_3},\cr
\CF_d^{g,1}&=\sum_{d_1,d_2,d_3} n^{g,1}_{d,d_1,d_2,d_3} q_1^{d_1} q_2^{d_2} q_3^{d_3/2},\cr
\CF_d^{g,2}&=\sum_{d_1,d_2,d_3} n^{g,2}_{d,d_1,d_2,d_3} q_1^{d_1} q_2^{d_2} q_3^{d_3},\cr
}}
where $d$ is the degree in $e^{-t}$, and the superscripts $g,c$ with $c=0,1,2$ denote the genus and the number of
crosscaps, respectively. Of course, $c=0$ is the oriented contribution obtained from 
\Ftopological\ (multiplied by the factor of $1/2$), and
in the second equation of \functionalsi\ $d_3$ must be odd. In order to compute these functionals,
we have to remove multicoverings according to \orcont\ and \strucun.
It is important to note that the requirement that the partition function satisfies the good integrality properties 
leading to \functionalsi\ is highly nontrivial.

\ifig\ptworptwo{Toric diagram for local ${\IP}^2$ attached to an $\IR\IP^2$}
{\epsfxsize1.9in\epsfbox{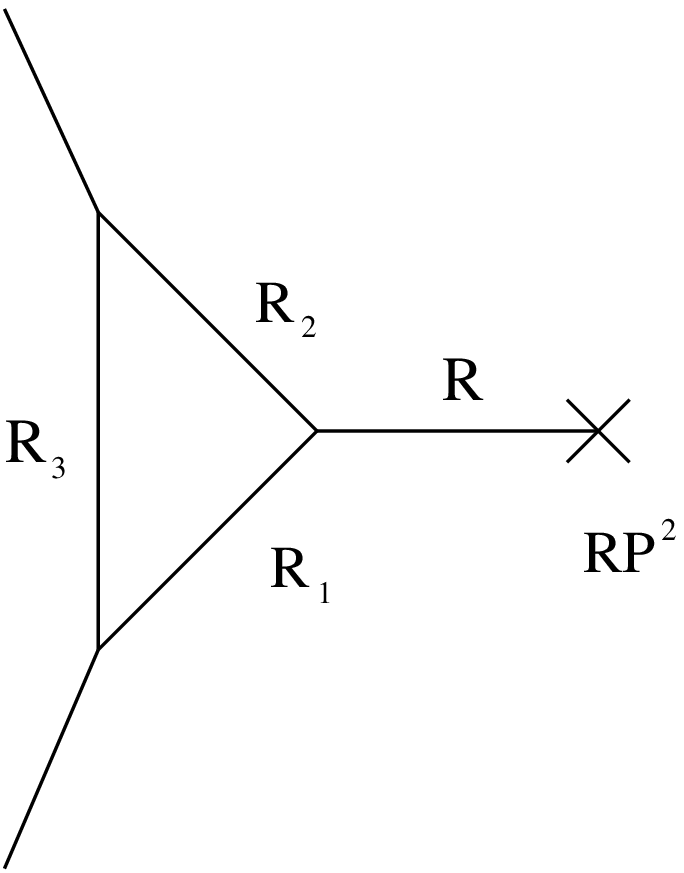}}

We present the results for the functionals given by \functionalsi\ in Appendix B. For the sake of brevity,
here we present the results only in the limiting case where we take the volumes of the two ${\IP}^1$'s attached
to the ${\IP}^2$ to infinity, as in \AMV. We thus obtain the answer for the simpler geometry whose toric diagram
is depicted in \ptworptwo. This geometry already captures all the interesting features of the unoriented and
oriented generating functionals.

To take this limit, one can directly consider the generating functionals \functionalsi\ and set $q_{1,2}=0$,
which corresponds to sending the two K\"ahler parameters to infinity. One can also obtain this limit by taking
the leading piece of the $U(N)$ Hopf link invariants in \Finstiv, in the sense explained in \leadinghopf. The
free energy of this geometry is just:

\eqn\prtwo{
\CF= -\ln \biggl\{  1+ \sum_{\ell} (-1)^{\ell} e^{-\ell t} q^{{1 \over 2}(\kappa_{R_1}+\kappa_{R_2}+\kappa_{R_3})}
q_3^{\ell(R_2)+\ell(R_3) \over 2} W_{R_3 R_1} W_{R_1 R_2} \CW_{R_2 R_3}^{SO(N)/Sp(N)}  \biggr\}.
}

The result can now be encoded in the simpler generating functionals

\eqn\functionalslimit{\eqalign{
\CF_d^{g,0}&={1 \over 2} \sum_{d_3} n^{g,0}_{d,d_3} q_3^{d_3},\cr
\CF_d^{g,1}&=\sum_{d_3} n^{g,1}_{d,d_3} q_3^{d_3/2},\cr
\CF_d^{g,2}&=\sum_{d_3} n^{g,2}_{d,d_3} q_3^{d_3},\cr
}}

\noindent with the same restrictions as for \functionalsi. We present here the all genus results we obtain up to degree
$6$ in $e^{-t}$. At this order $n_{d,d_3}^{g,c}=0~\forall~g \geq 11$ (all the invariants $n_{d,d_3}^{g,c}$ with $d \leq 6$ 
that are not shown in the tables are understood to be zero). The results in Tables $1-16$ correspond to $Sp(N)$ gauge group; 
to obtain the $SO(N)$ result it suffices to change the sign of the invariants with $c=1$.

{\vbox{\sevenpoint{
$$
\vbox{\offinterlineskip\tabskip=0pt
\halign{\strut
\vrule#
&
&\hfil ~$#$ \vrule
&\hfil ~$#$
&\hfil ~$#$
&\hfil ~$#$
&\hfil ~$#$
&\hfil ~$#$
&\hfil ~$#$
&\vrule \hfil ~$#$ \vrule
&\hfil ~$#$
&\hfil ~$#$
&\hfil ~$#$
&\hfil ~$#$
&\hfil ~$#$
&\hfil ~$#$
&\vrule #\cr
\noalign{\hrule}
&c=0
&d_3=0
&1
&2
&3
&4
&5
&c=1
&d_3=1
&3
&5
&7
&9
&11
&
\cr
\noalign{\hrule}
&d=0
&0
&1
&0
&0
&0
&0
&d=0
&1
&0
&0
&0
&0
&0
&
\cr
&1
&6
&-4
&0
&0
&0
&0
&1
&-2
&0
&0
&0
&0
&0
&
\cr
&2
&-12
&14
&-2
&0
&0
&0
&2
&5
&-3
&0
&0
&0
&0
&
\cr
&3
&54
&-84
&30
&0
&0
&0
&3
&-32
&30
&-4
&0
&0
&0
&
\cr
&4
&-384
&725
&-392
&51
&0
&0
&4
&286
&-369
&112
&-5
&0
&0
&
\cr
&5
&3390
&-7540
&5434
&-1368
&84
&0
&5
&-3038
&5016
&-2410
&328
&-6
&0
&
\cr
&6
&-34128
&87776
&-79198
&29466
&-4040
&124
&6
&35870
&-72150
&47554
&-11528
&819
&-7
&
\cr
}\hrule}$$}
\vskip - 7 mm
\centerline{\ninepoint {\bf Table 1:} Invariants $n_{d,d_3}^{0,c}$ at genus 0, with $c=0,1$, up to $d=6$.}
\vskip7pt}
\noindent
\smallskip

{\vbox{\ninepoint{
$$
\vbox{\offinterlineskip\tabskip=0pt
\halign{\strut
\vrule#
&
&\hfil ~$#$ \vrule
&\hfil ~$#$
&\hfil ~$#$
&\hfil ~$#$
&\hfil ~$#$
&\vrule #\cr
\noalign{\hrule}
&c=2
&d_3=2
&3
&4
&5
&
\cr
\noalign{\hrule}
&d=3
&1
&0
&0
&0
&\cr
&4
&-11
&2
&0
&0
&
\cr
&5
&131
&-66
&7
&0
&
\cr
&6
&-1690
&1460
&-333
&12
&
\cr
}\hrule}$$}
\vskip - 7 mm
\centerline{{\bf Table 2:} Invariants $n_{d,d_3}^{0,2}$ at genus 0, up to $d=6$.}
\vskip7pt}
\noindent
\smallskip

{\vbox{\ninepoint{
$$
\vbox{\offinterlineskip\tabskip=0pt
\halign{\strut
\vrule#
&
&\hfil ~$#$ \vrule
&\hfil ~$#$
&\hfil ~$#$
&\hfil ~$#$
&\hfil ~$#$
&\hfil ~$#$
&\vrule \hfil ~$#$ \vrule
&\hfil ~$#$
&\hfil ~$#$
&\hfil ~$#$
&\hfil ~$#$
&\hfil ~$#$
&\vrule #\cr
\noalign{\hrule}

&c=0
&d_3=0
&1
&2
&3
&4
&c=1
&d_3=1
&3
&5
&7
&9
&
\cr
\noalign{\hrule}

&d=3
&20
&-18
&0
&0
&0
&d=3
&-9
&7
&0
&0
&0
&
\cr
&4
&-462
&612
&-168
&0
&0
&4
&288
&-315
&59
&0
&0
&
\cr
&5
&8904
&-15210
&7380
&-930
&0
&5
&-6984
&9954
&-3630
&282
&0
&
\cr
&6
&-161896
&336636
&-228532
&56536
&-3851
&6
&152622
&-269501
&145467
&-25672
&1014
&
\cr
}\hrule}$$}
\vskip - 7 mm
\centerline{{\bf Table 3:} Invariants $n_{d,d_3}^{1,c}$ at genus 1, with $c=0,1$, up to $d=6$.}
\vskip7pt}
\noindent
\smallskip

{\vbox{\ninepoint{
$$
\vbox{\offinterlineskip\tabskip=0pt
\halign{\strut
\vrule#
&
&\hfil ~$#$ \vrule
&\hfil ~$#$
&\hfil ~$#$
&\hfil ~$#$
&\hfil ~$#$
&\vrule #\cr
\noalign{\hrule}
&c=2
&d_3=2
&3
&4
&5
&
\cr
\noalign{\hrule}
&d=4
&-6
&0
&0
&0
&
\cr
&5
&201
&-55
&1
&0
&
\cr
&6
&-5180
&3180
&-438
&2
&
\cr
}\hrule}$$}
\vskip - 7 mm
\centerline{{\bf Table 4:} Invariants $n_{d,d_3}^{1,2}$ at genus 1, up to $d=6$.}
\vskip7pt}
\noindent
\smallskip

{\vbox{\ninepoint{
$$
\vbox{\offinterlineskip\tabskip=0pt
\halign{\strut
\vrule#
&
&\hfil ~$#$ \vrule
&\hfil ~$#$
&\hfil ~$#$
&\hfil ~$#$
&\hfil ~$#$
&\hfil ~$#$
&\vrule \hfil ~$#$ \vrule
&\hfil ~$#$
&\hfil ~$#$
&\hfil ~$#$
&\hfil ~$#$
&\hfil ~$#$
&\vrule #\cr
\noalign{\hrule}

&c=0
&d_3=0
&1
&2
&3
&4
&c=1
&d_3=1
&3
&5
&7
&9
&
\cr
\noalign{\hrule}

&d=4
&-204
&216
&-24
&0
&0
&d=4
&108
&-103
&9
&0
&0
&
\cr
&5
&10860
&-15444
&5154
&-276
&0
&5
&-7506
&9474
&-2567
&95
&0
&
\cr
&6
&-388044
&690273
&-365536
&60235
&-1800
&6
&329544
&-521400
&231550
&-29010
&554
&
\cr
}\hrule}$$}
\vskip - 7 mm
\centerline{{\bf Table 5:} Invariants $n_{d,d_3}^{2,c}$ at genus 2, with $c=0,1$ up to $d=6$.}
\vskip7pt}
\noindent
\smallskip

{\vbox{\ninepoint{
$$
\vbox{\offinterlineskip\tabskip=0pt
\halign{\strut
\vrule#
&
&\hfil ~$#$ \vrule
&\hfil ~$#$
&\hfil ~$#$
&\hfil ~$#$
&\vrule #\cr
\noalign{\hrule}
&c=2
&d_3=2
&3
&4
&
\cr
\noalign{\hrule}
&d=4
&-1
&0
&0
&
\cr
&5
&146
&-18
&0
&
\cr
&6
&-8296
&3520
&-274
&
\cr
}\hrule}$$}
\vskip - 7 mm
\centerline{{\bf Table 6:} Invariants $n_{d,d_3}^{2,2}$ at genus 2, up to $d=6$.}
\vskip7pt}
\noindent
\smallskip

{\vbox{\ninepoint{
$$
\vbox{\offinterlineskip\tabskip=0pt
\halign{\strut
\vrule#
&
&\hfil ~$#$ \vrule
&\hfil ~$#$
&\hfil ~$#$
&\hfil ~$#$
&\hfil ~$#$
&\hfil ~$#$
&\vrule \hfil ~$#$ \vrule
&\hfil ~$#$
&\hfil ~$#$
&\hfil ~$#$
&\hfil ~$#$
&\hfil ~$#$
&\vrule #\cr
\noalign{\hrule}

&c=0
&d_3=0
&1
&2
&3
&4
&c=1
&d_3=1
&3
&5
&7
&9
&
\cr
\noalign{\hrule}

&d=4
&-30
&28
&0
&0
&0
&d=4
&14
&-12
&0
&0
&0
&
\cr
&5
&7344
&-9094
&2036
&-30
&0
&5
&-4519
&5133
&-977
&11
&0
&
\cr
&6
&-581706
&913220
&-381934
&40728
&-408
&6
&447502
&-642780
&233460
&-19781
&139
&
\cr
}\hrule}$$}
\vskip - 7 mm
\centerline{{\bf Table 7:} Invariants $n_{d,d_3}^{3,c}$ at genus 3, with $c=0,1$, up to $d=6$.}
\vskip7pt}
\noindent
\smallskip

{\vbox{\ninepoint{
$$
\vbox{\offinterlineskip\tabskip=0pt
\halign{\strut
\vrule#
&
&\hfil ~$#$ \vrule
&\hfil ~$#$
&\hfil ~$#$
&\hfil ~$#$
&\vrule #\cr
\noalign{\hrule}
&c=2
&d_3=2
&3
&4
&
\cr
\noalign{\hrule}
&d=5
&58
&-2
&0
&
\cr
&6
&-8489
&2352
&-90
&
\cr
}\hrule}$$}
\vskip - 7 mm
\centerline{{\bf Table 8:} Invariants $n_{d,d_3}^{3,2}$ at genus 3, up to $d=6$.}
\vskip7pt}
\noindent
\smallskip

{\vbox{\ninepoint{
$$
\vbox{\offinterlineskip\tabskip=0pt
\halign{\strut
\vrule#
&
&\hfil ~$#$ \vrule
&\hfil ~$#$
&\hfil ~$#$
&\hfil ~$#$
&\hfil ~$#$
&\hfil ~$#$
&\vrule \hfil ~$#$ \vrule
&\hfil ~$#$
&\hfil ~$#$
&\hfil ~$#$
&\hfil ~$#$
&\hfil ~$#$
&\vrule #\cr
\noalign{\hrule}

&c=0
&d_3=0
&1
&2
&3
&4
&c=1
&d_3=1
&3
&5
&7
&9
&
\cr
\noalign{\hrule}

&d=5
&2772
&-3084
&424
&0
&0
&d=5
&-1542
&1599
&-191
&0
&0
&
\cr
&6
&-580800
&821490
&-270708
&17600
&-36
&6
&407661
&-536973
&157255
&-8372
&13
&
\cr
}\hrule}$$}
\vskip - 7 mm
\centerline{{\bf Table 9:} Invariants $n_{d,d_3}^{4,c}$ at genus 4, with $c=0,1$, up to $d=6$.}
\vskip7pt}
\noindent
\smallskip

{\vbox{\ninepoint{
$$
\vbox{\offinterlineskip\tabskip=0pt
\halign{\strut
\vrule#
&
&\hfil ~$#$ \vrule
&\hfil ~$#$
&\hfil ~$#$
&\hfil ~$#$
&\vrule #\cr
\noalign{\hrule}
&c=2
&d_3=2
&3
&4
&
\cr
\noalign{\hrule}
&d=5
&12
&0
&0
&
\cr
&6
&-5862
&976
&-15
&
\cr
}\hrule}$$}
\vskip - 7 mm
\centerline{{\bf Table 10:} Invariants $n_{d,d_3}^{4,2}$ at genus 4, up to $d=6$.}
\vskip7pt}
\noindent
\smallskip

{\vbox{\sevenpoint{
$$
\vbox{\offinterlineskip\tabskip=0pt
\halign{\strut
\vrule#
&
&\hfil ~$#$ \vrule
&\hfil ~$#$
&\hfil ~$#$
&\hfil ~$#$
&\hfil ~$#$
&\vrule \hfil ~$#$ \vrule
&\hfil ~$#$
&\hfil ~$#$
&\hfil ~$#$
&\hfil ~$#$
&\vrule \hfil ~$#$ \vrule
&\hfil ~$#$
&\hfil ~$#$
&\hfil ~$#$
&\vrule #\cr
\noalign{\hrule}

&c=0
&d_3=0
&1
&2
&3
&c=1
&d_3=1
&3
&5
&7
&c=2
&d_3=2
&3
&4
&
\cr
\noalign{\hrule}

&d=5
&540
&-552
&36
&0
&d=5
&-276
&265
&-15
&0
&d=5
&1
&0
&0
&
\cr
&6
&-393714
&509724
&-130496
&4684
&6
&254310
&-309962
&71523
&-2141
&6
&-2758
&245
&-1
&
\cr
}\hrule}$$}
\vskip - 7 mm
\centerline{\ninepoint {\bf Table 11:} Invariants $n_{d,d_3}^{5,c}$ at genus 5, up to $d=6$.}
\vskip7pt}
\noindent
\smallskip

{\vbox{\ninepoint{
$$
\vbox{\offinterlineskip\tabskip=0pt
\halign{\strut
\vrule#
&
&\hfil ~$#$ \vrule
&\hfil ~$#$
&\hfil ~$#$
&\hfil ~$#$
&\hfil ~$#$
&\vrule \hfil ~$#$ \vrule
&\hfil ~$#$
&\hfil ~$#$
&\hfil ~$#$
&\hfil ~$#$
&\vrule \hfil ~$#$ \vrule
&\hfil ~$#$
&\hfil ~$#$
&\vrule #\cr
\noalign{\hrule}

&c=0
&d_3=0
&1
&2
&3
&c=1
&d_3=1
&3
&5
&7
&c=2
&d_3=2
&3
&
\cr
\noalign{\hrule}

&d=5
&42
&-40
&0
&0
&d=5
&-20
&18
&0
&0
&d=5
&0
&0
&
\cr
&6
&-180780
&216960
&-41904
&696
&6
&108440
&-123342
&21630
&-302
&6
&-868
&34
&
\cr
}\hrule}$$}
\vskip - 7 mm
\centerline{{\bf Table 12:} Invariants $n_{d,d_3}^{6,c}$ at genus 6, up to $d=6$.}
\vskip7pt}
\noindent
\smallskip

{\vbox{\ninepoint{
$$
\vbox{\offinterlineskip\tabskip=0pt
\halign{\strut
\vrule#
&
&\hfil ~$#$ \vrule
&\hfil ~$#$
&\hfil ~$#$
&\hfil ~$#$
&\hfil ~$#$
&\vrule \hfil ~$#$ \vrule
&\hfil ~$#$
&\hfil ~$#$
&\hfil ~$#$
&\hfil ~$#$
&\vrule \hfil ~$#$ \vrule
&\hfil ~$#$
&\hfil ~$#$
&\vrule #\cr
\noalign{\hrule}

&c=0
&d_3=0
&1
&2
&3
&c=1
&d_3=1
&3
&5
&7
&c=2
&d_3=2
&3
&
\cr
\noalign{\hrule}

&d=6
&-55076
&61896
&-8532
&44
&d=6
&30948
&-33110
&4156
&-18
&d=6
&-174
&2
&

\cr
}\hrule}$$}
\vskip - 7 mm
\centerline{{\bf Table 13:} Invariants $n_{d,d_3}^{7,c}$ at genus 7, up to $d=6$.}
\vskip7pt}
\noindent
\smallskip

{\vbox{\ninepoint{
$$
\vbox{\offinterlineskip\tabskip=0pt
\halign{\strut
\vrule#
&
&\hfil ~$#$ \vrule
&\hfil ~$#$
&\hfil ~$#$
&\hfil ~$#$
&\vrule \hfil ~$#$ \vrule
&\hfil ~$#$
&\hfil ~$#$
&\hfil ~$#$
&\vrule \hfil ~$#$ \vrule
&\hfil ~$#$
&\vrule #\cr
\noalign{\hrule}

&c=0
&d_3=0
&1
&2
&c=1
&d_3=1
&3
&5
&c=2
&d_3=2
&
\cr
\noalign{\hrule}

&d=6
&-10620
&11268
&-992
&d=6
&5634
&-5710
&458
&d=6
&-20
&

\cr
}\hrule}$$}
\vskip - 7 mm
\centerline{{\bf Table 14:} Invariants $n_{d,d_3}^{8,c}$ at genus 8, up to $d=6$.}
\vskip7pt}
\noindent
\smallskip

{\vbox{\ninepoint{
$$
\vbox{\offinterlineskip\tabskip=0pt
\halign{\strut
\vrule#
&
&\hfil ~$#$ \vrule
&\hfil ~$#$
&\hfil ~$#$
&\hfil ~$#$
&\vrule \hfil ~$#$ \vrule
&\hfil ~$#$
&\hfil ~$#$
&\hfil ~$#$
&\vrule \hfil ~$#$ \vrule
&\hfil ~$#$
&\vrule #\cr
\noalign{\hrule}

&c=0
&d_3=0
&1
&2
&c=1
&d_3=1
&3
&5
&c=2
&d_3=2
&
\cr
\noalign{\hrule}

&d=6
&-1170
&1180
&-50
&d=6
&590
&-570
&22
&d=6
&-1
&

\cr
}\hrule}$$}
\vskip - 7 mm
\centerline{{\bf Table 15:} Invariants $n_{d,d_3}^{9,c}$ at genus 9, up to $d=6$.}
\vskip7pt}
\noindent
\smallskip

{\vbox{\ninepoint{
$$
\vbox{\offinterlineskip\tabskip=0pt
\halign{\strut
\vrule#
&
&\hfil ~$#$ \vrule
&\hfil ~$#$
&\hfil ~$#$
&\vrule \hfil ~$#$ \vrule
&\hfil ~$#$
&\hfil ~$#$
&\vrule #\cr
\noalign{\hrule}

&c=0
&d_3=0
&1
&c=1
&d_3=1
&3
&
\cr
\noalign{\hrule}

&d=6
&-56
&54
&d=6
&27
&-25
&

\cr
}\hrule}$$}
\vskip - 7 mm
\centerline{{\bf Table 16:} Invariants $n_{d,d_3}^{10,c}$ at genus 10, up to $d=6$.}
\vskip7pt}
\noindent
\smallskip

\newsec{Unoriented Localization}

As explained in section $2$, to compute the full partition function of closed topological strings on the geometry 
before the geometric transition, we have to sum both over holomorphic maps from orientable Riemann surfaces to the 
Calabi-Yau space $X$ as well as maps from non-orientable worldsheets to the orientifolded space $X/I$.

In \DFM\ it was developed a method for summing unoriented world-sheet instantons for
closed topological strings based on localization with respect to a torus action on a moduli space of equivariant holomorphic 
maps. Although in \DFM\ this moduli space has not been constructed, a computational definition for its virtual fundamental cycle 
was given. Concretely, this reduces to enumerating all fixed loci under an induced torus action on the moduli space and 
assigning a local contribution to each component of the fixed locus using an equivariant version of the localization 
theorem of \GP. Moreover, in \DFM\ it was shown that the fixed loci can be represented in terms of Kontsevich graphs \Ko\ 
with involution. 

This method does not rely on large $N$ duality, and therefore may provide an independent check of our large $N$ duality proposal 
for orientifolds. Namely, we can employ the localization techniques of \DFM\ to compute one crosscap and two crosscaps contributions to the
full closed topological string partition function on the orientifolded geometry before the geometric transition.

We can use the computation in \DFM\ to confirm the one crosscap invariants for low degree and genus obtained 
from the Chern-Simons computation. There, it was computed the unoriented free
energy for a ${\IP^2}$ with a ${\IR\IP^2}$
attached. This is exactly the limiting geometry for which we presented our results in Tables $1-16$, related to the full 
geometry of section $2$ by sending the two K\"ahler parameters of the two ${\IP^1}$'s of the full geometry to infinity. 
In our variables, the result of \DFM\ reads 
\eqn\localization{\eqalign{
\CF=&{1 \over g_s} (q_3^{1/2}-2 e^{-t} q_3^{1/2} + 5 e^{-2t} q_3^{1/2}+...+{1 \over 9} q_3^{3/2} -
3 e^{-2t} q_3^{3/2}+{268 \over 9} e^{-3t} q_3^{3/2}+...)\cr
&+g_s ({1 \over 24} q_3^{1/2}-{1\over 12}e^{-t} q_3^{1/2} + ...).
}}
By expanding $q=e^{g_s}$ in powers of $g_s$, it is straightforward to show that the contributions with $c=1$ 
in Tables $1-16$ are in agreement with \localization.

In the following we will compute some Klein bottle amplitudes using unoriented localization.
We will find agreement with the Chern-Simons and with the topological vertex computations presented in the 
next section. We will perform the computations in the Calabi-Yau geometry $\widetilde X$. In the patch $\{X_1\neq 0,
X_7\neq 0,X_{10}\neq 0\}$
we introduce local coordinates
\eqn\locor{
z={X_1^3X_4\over X_7X_{10}^3},\quad u={X_6X_7X_{10}^2\over X_1},\quad v={X_5X_7X_{10}^2\over X_1}.
}
Using \constr\ we obtain the weights of the local coordinates
\eqn\locw{
\l_z=6\l_1+2\l_4,\quad\l_u=-3\l_1-\l_4+\l_6,\quad\l_v=-3\l_1-\l_4+\l_5.
}
Note that the compatibility of the involution with the torus action implies $\l_z+\l_u+\l_v=0$.

We will denote the contributions of the fixed loci by $C_{\chi,d,h}$, where $\chi$ is the Euler characteristic of the 
unoriented source Riemann surface and $d$ and $h$ are the degrees of the map with respect to the $\IR\IP^2$ and hyperplane 
class of $\IP^2$ respectively.

\subsec{Unoriented localization $@$ $2$ crosscaps $@$ degree $2$ $\IR\IP^2$}

The computation at degree $0$ hyperplane class has been performed in \DFM . Let us recall the graphs
and their contributions.

\ifig\unverz{Two crosscaps and no hyperplane at degree $2$ $\IR\IP^2$.}
{\epsfxsize2.5in\epsfbox{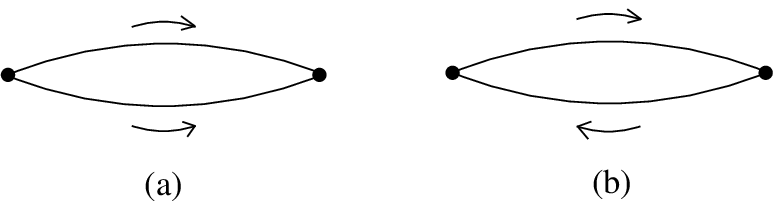}}

Note that in case $(b)$ the antiholomorphic involution exchanges the two components of the source curve. In
\DFM\ it has been postulated that such an operation will introduce an additional minus sign. Therefore the
contributions of the two graphs are
\eqn\tcnh{
C_{0,2,0}^{(a)}={\l_u\l_v\over 4\l_z^2},\qquad C_{0,2,0}^{(b)}=-{\l_u\l_v\over 4\l_z^2}.
}

Let us consider now the degree $1$ hyperplane class configurations. The graphs allowed are presented in fig. 9
below.

\ifig\unvervb{Two crosscaps and one hyperplane at degree $2$ $\IR\IP^2$. Mirror pairs are $\{(a),(c)\}$ and $\{(b),(d)\}$
respectively.}
{\epsfxsize3.0in\epsfbox{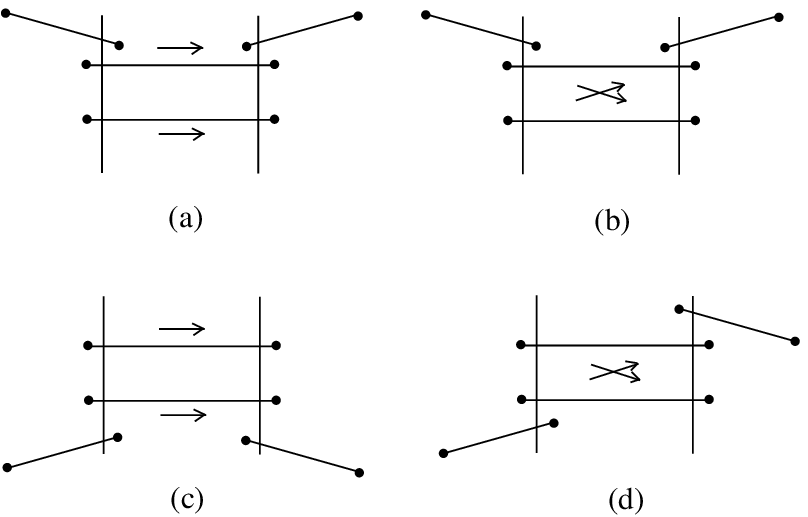}}

The allowed configurations are obtained by performing bubblings at the nodes of the graphs in \unverz\ and
inserting degree $1$ hyperplane graphs; we will call such configurations type ${\rm I}$ graphs. These come in pairs, each one 
admits a mirror graph. From now on, we will draw a single graph for each mirror pair. The contributions of the above configurations 
are given by
\eqn\consdego{
C_{0,2,1}^{(a)}={\l_v^2\over 2\l_z^2},\quad  C_{0,2,1}^{(b)}=-{\l_v^2\over 2\l_z^2},
\quad C_{0,2,1}^{(c)}={\l_u^2\over 2\l_z^2},\quad  C_{0,2,1}^{(d)}=-{\l_u^2\over 2\l_z^2}
}
where we have used again the sign rule postulated in \DFM. The graph contributions add up to zero.

The discussion is similar at degree $2$ hyperplane class. The type ${\rm I}$ graphs appearing cancel in pairs due to the same sign rule 
as above. There also appear new configurations, which we will call type ${\rm II}$ graphs, and which we present in fig. $10$ below.  

\ifig\unverbi{Two crosscaps and two hyperplanes at degree $2$ $\IR\IP^2$: type ${\rm II}$ graphs.}
{\epsfxsize3.2in\epsfbox{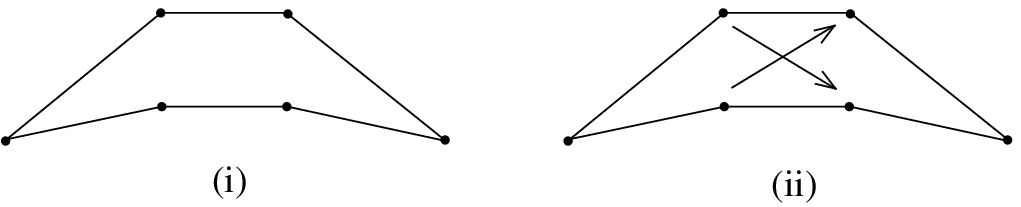}}

\noindent Their contributions are given by 
\eqn\consdegoi{
C_{0,2,2}^{(i)}={(\l_u-2\l_v)(\l_v-\l_u)\over 2\l_v^2},\quad C_{0,2,2}^{(ii)}=-{(\l_u-2\l_v)(\l_v-\l_u)\over 2\l_v^2}
}
and therefore they cancel due to the same sign rule that we used previously.

At degree $3$ hyperplane class, we obtain again pairs of graphs of type ${\rm I}$ and type ${\rm II}$ that
cancel each other. In fig. $11$ we draw some new type ${\rm II}$ graphs whose analogues at higher $\IR\IP^2$ degree will play an important 
role.

\ifig\unver{Two crosscaps and three hyperplanes at degree $2$ $\IR\IP^2$: type ${\rm II}$ graphs.}
{\epsfxsize3.2in\epsfbox{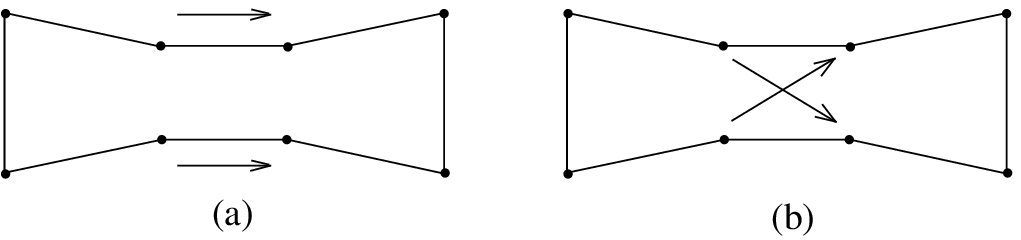}}

The contributions of the above two graphs are: $C^{(a)}_{0,2,3}=-C^{(b)}_{0,2,3}=1$. To conclude,
we obtain that up to degree $3$ hyperplane class, the $2$ crosscaps degree $2$ $\IR\IP^2$ Gromov-Witten invariants vanish. 
In fact, this will be true at any hyperplane class degree.

\subsec{Unoriented localization $@$ $2$ crosscaps $@$ degree $4$ $\IR\IP^2$}

At degree $0$ hyperplane class this computation has been performed in \DFM . We list the graphs

\ifig\unver{Two crosscaps and no hyperplane at degree $4$ $\IR\IP^2$.}
{\epsfxsize5.2in\epsfbox{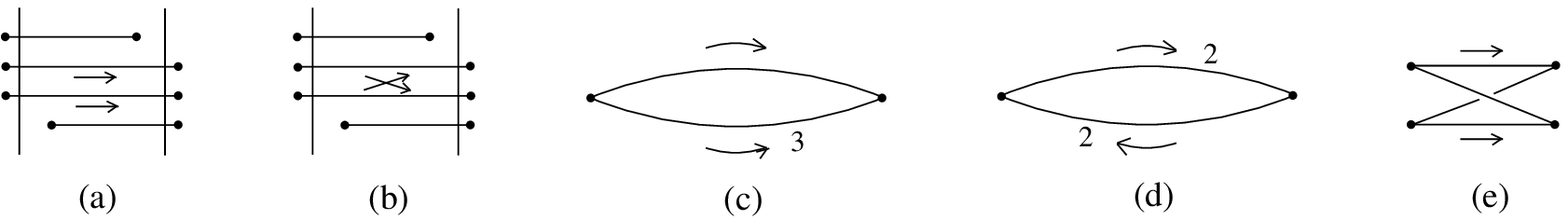}}

\noindent and their contributions
\eqn\nohyp{\eqalign{
& C_{0,4,0}^{(a)}={1\over 2}{\l_u^2\l_v^2\over \l_z^4}, \qquad C_{0,4,0}^{(b)}=-{1\over 2}{\l_u^2\l_v^2\over \l_z^4},
\qquad C_{0,4,0}^{(c)}=
{1\over 8}{\l_u\l_v(2\l_z^2-9\l_u\l_v)\over \l_z^4}, \cr
& C_{0,4,0}^{(d)}=-{1\over 4}{\l_u\l_v(\l_z^2-4\l_u\l_v)\over \l_z^4},
\qquad C_{0,4,0}^{(e)}={1\over 8}{\l_u^2\l_v^2\over \l_z^4}.}}
Note that $C_{0,4,0}^{(a)}+C_{0,4,0}^{(b)}=0$ and $C_{0,4,0}^{(c)}+C_{0,4,0}^{(d)}+C_{0,4,0}^{(e)}=0$.

At degree $1$ hyperplane class there appear new configurations, which we will call type ${\rm III}$ graphs; they are obtained by adding 
to the first two graphs in \unver\ degree $1$ hyperplane lines as shown below.

\ifig\unverii{Two crosscaps and one hyperplane at degree $4$ $\IR\IP^2$: type ${\rm III}$ graphs.}
{\epsfxsize3.7in\epsfbox{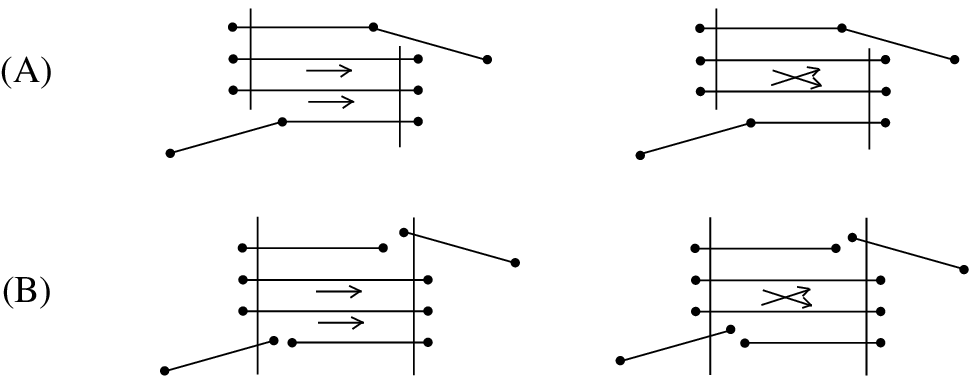}}

Using again the sign rule in \DFM,\ the two graphs in each line of \unverii\ add up to zero. We now turn 
to type ${\rm I}$ graphs; they
are presented in fig. $14$ and their contributions are:

\ifig\unveriii{Two crosscaps and one hyperplane at degree $4$ $\IR\IP^2$: type ${\rm I}$ graphs.}
{\epsfxsize5.0in\epsfbox{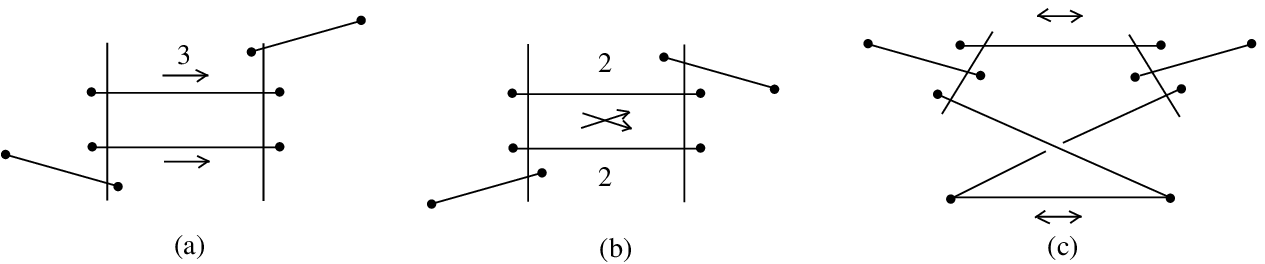}}

$$
C_{0,4,1}^{(a)}={1\over 2}(2-9{\l_u\l_v\over\l_z^2})({\l_u^2\over\l_z^2}+{\l_v^2\over\l_z^2}),~~
C_{0,4,1}^{(b)}=(-1+4{\l_u\l_v\over\l_z^2})({\l_u^2\over\l_z^2}+{\l_v^2\over\l_z^2}),~~
C_{0,4,1}^{(c)}={\l_u\l_v\over 2\l_z^2}({\l_u^2\over\l_z^2}+{\l_v^2\over\l_z^2}).
$$
It is easy to check that $C_{0,4,1}^{(a)}+C_{0,4,1}^{(b)}+C_{0,4,1}^{(c)}=0$. This is in fact the same
cancellation that took place at degree $0$ hyperplane class between the contributions of the corresponding
three graphs. Again, we see that at degree $1$ hyperplane class there is nothing essentially new compared to
degree $0$ hyperplane class.

Let us now consider the case of degree $2$ hyperplane class. We can split the allowed configurations in graphs
of type ${\rm I}$ and ${\rm III}$ above. Configurations of type ${\rm III}$ are built by starting with the graphs $(a)$ and $(b)$
in \unver\ and further adding in all possible ways degree $2$ graphs in $\IP^2$. They
will always cancel in pairs. Configurations of type ${\rm I}$ are constructed by starting with
the graphs $(c)$, $(d)$ and $(e)$ in \unver,\ performing a bubbling at a pair of identified nodes and inserting degree $2$ graphs
in $\IP^2$. The contributions of the graphs with degree $2$ multicoverings of one of the hyperplane sections cancel as before; there also 
appear configurations as in fig. $15$ below. However, their contributions also add up to zero, and this will be true for any quartet of 
type ${\rm I}$ graphs as in fig. $15$. 

\ifig\unveriv{Two crosscaps and one hyperplane at degree $4$ $\IR\IP^2$: type ${\rm I}$ graphs.}
{\epsfxsize6.0in\epsfbox{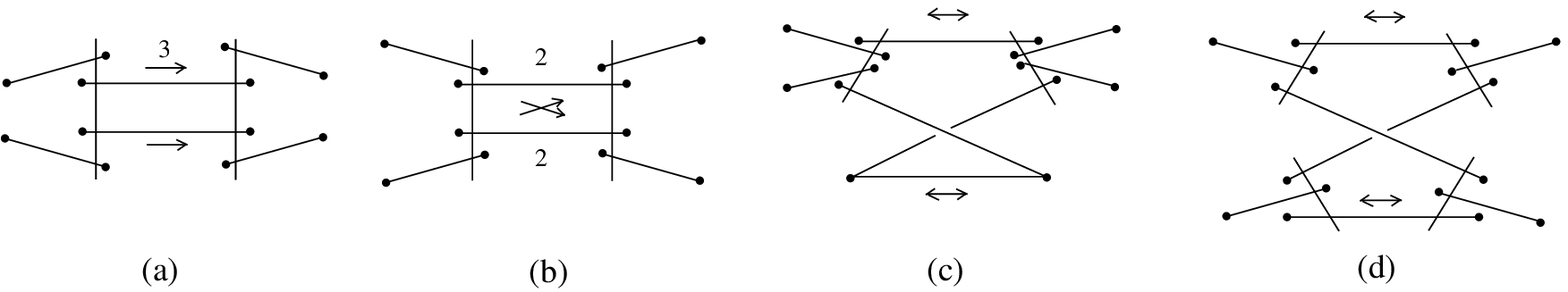}}

There are also type ${\rm II}$ graphs, which are constructed by starting with the graphs $(c)$, $(d)$ and $(e)$ in \unver. A triplet 
of such graphs is presented below.

\ifig\unvervi{Two crosscaps and two hyperplanes at degree $4$ $\IR\IP^2$: type ${\rm II}$ graphs.}
{\epsfxsize5.0in\epsfbox{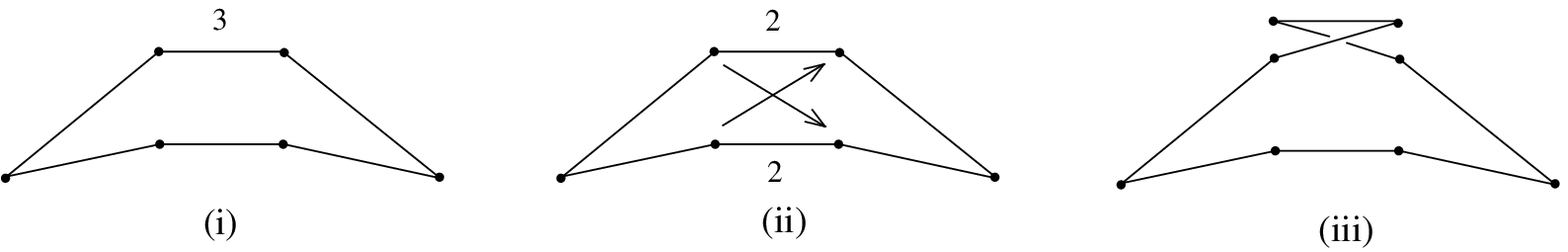}}

\noindent The total contribution of the above three graphs is 
\eqn\newtypeii{
C_{0,4,2}^{(i)}+C_{0,4,2}^{(ii)}+C_{0,4,2}^{(iii)}=-{(\l_u-\l_v)^2(\l_u+\l_v+\l_z)Q_9(\l_u,\l_v,\l_z)\over 4
\l_u^2\l_v^2\l_z^2(2\l_u+\l_z)^2(2\l_v+\l_z)^2(3\l_u+\l_z)(3\l_v+\l_z)},
}
where $Q_9(\l_u,\l_v,\l_z)$ is a degree $9$ homogeneous polynomial in $\l_u,\l_v,\l_z$. We recall that consistency 
of the antiholomorphic involution 
with the torus action implies $\l_u+\l_v+\l_z=0$, and therefore the sum of the graphs in \unvervi\ is zero. This will also be true 
for the other possible triplet of type ${\rm II}$ graphs. 
We conclude that up to degree $2$ hyperplane class, the two crosscaps degree $4$ $\IR\IP^2$ Gromov-Witten invariants vanish.

At degree $3$ hyperplane class there appear all three types of graphs. We claim that the 
type ${\rm I}$ and ${\rm III}$ graphs sum up to zero, as above. Besides sets of type ${\rm II}$ graphs that have analogues at lower 
degree hyperplane class, and whose contributions add up to zero in a similar fashion, at degree $3$ hyperplane class there also are new 
collections of graphs. Such a set is presented in fig. $17$.

\ifig\unveriVV{Two crosscaps and three hyperplanes at degree $4$ $\IR\IP^2$: type ${\rm II}$ graphs.}
{\epsfxsize5.0in\epsfbox{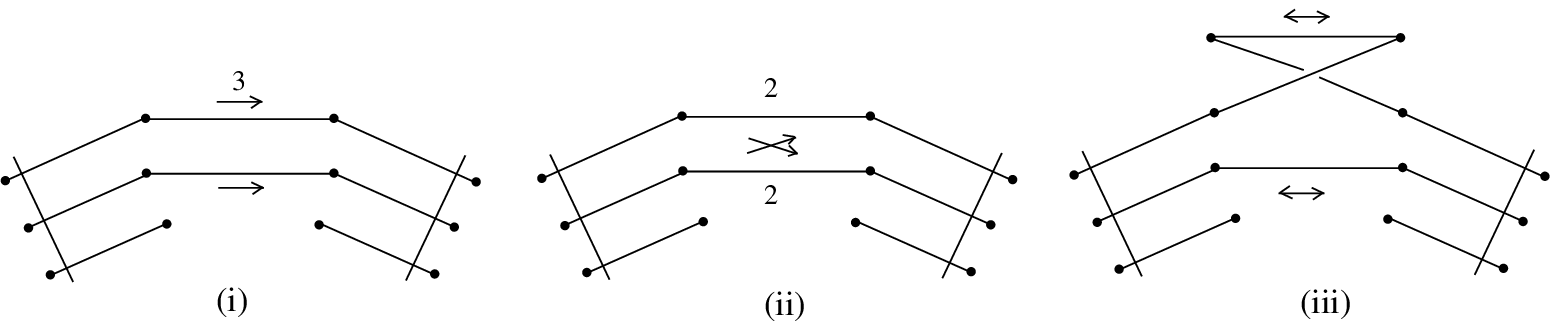}}
\noindent The total contribution of the above three graphs is 
\eqn\newtypeii{
C_{0,4,3}^{(i)}+C_{0,4,3}^{(ii)}+C_{0,4,3}^{(iii)}={\l_u(\l_u+\l_v+\l_z)(\l_u-\l_v)^2(\l_u-2\l_v)^2Q_3(\l_u,\l_v,\l_z)\over 2
\l_z^2\l_v^4(2\l_v+\l_z)^2(3\l_v+\l_z)},
}
where $Q_3(\l_u,\l_v,\l_z)$ is a degree $3$ homogeneous polynomial in $\l_u,\l_v,\l_z$. But $\l_u+\l_v+\l_z=0$, and these graphs 
sum up to zero.

However, at degree $3$ hyperplane class there is a unique set of type ${\rm II}$ graphs, presented in fig. $18$, whose total 
contribution does not vanish.

\ifig\unveriV{Two crosscaps and three hyperplanes at degree $4$ $\IR\IP^2$: type ${\rm II}$ graphs.}
{\epsfxsize5.0in\epsfbox{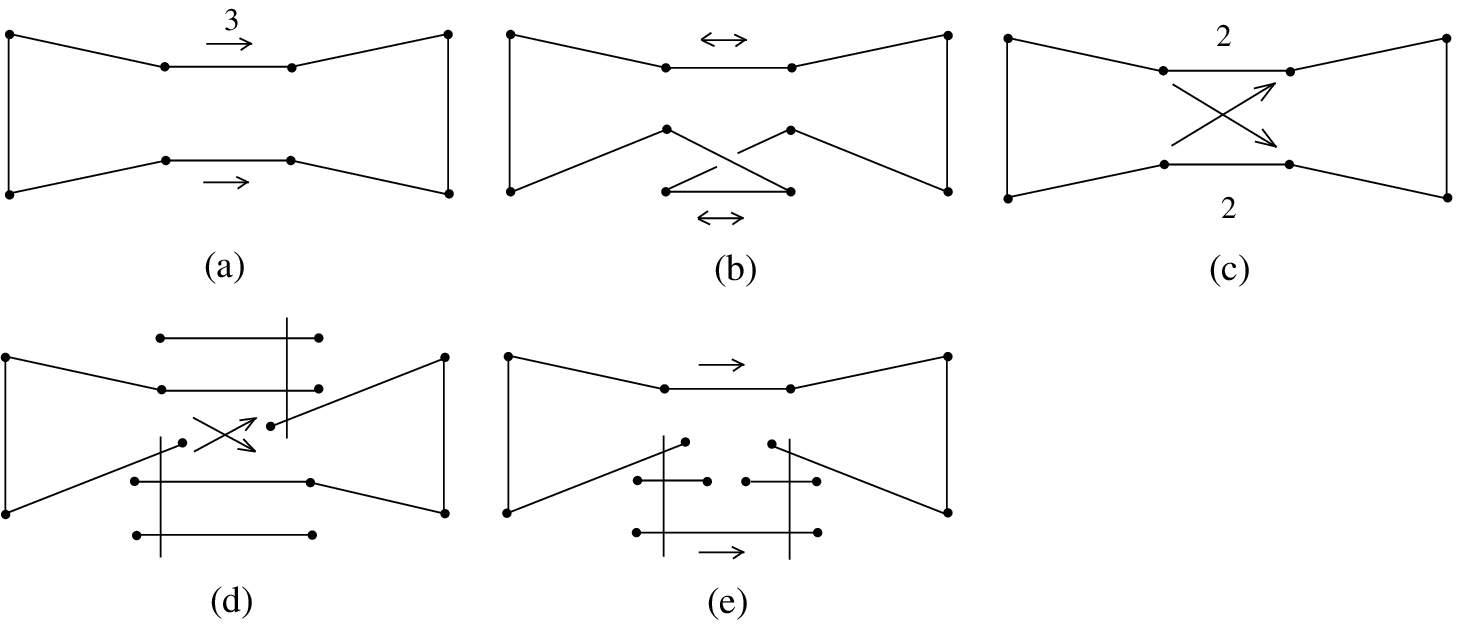}}

\noindent The contributions of the graphs in \unveriV\ are given by
\eqn\disapp{\eqalign{
&C_{0,4,3}^{(a)}=1-3{\l_u\l_v\over\l_z^2},\quad C_{0,4,3}^{(b)}={\l_u\l_v\over\l_z^2},\quad C_{0,4,3}^{(c)}=2{\l_u\l_v\over\l_z^2},\cr
&C_{0,4,3}^{(d)}=-1+2{\l_u\l_v\over\l_z^2},\quad C_{0,4,3}^{(e)}=1-2{\l_u\l_v\over\l_z^2}.
}}

We see that the sum of the above expressions is equal to $1$, which is the Gromov-Witten invariant $n_{3,2}^{0,2}$ of Table 1. It is 
straightforward to perform now a similar computation but taking also into account the two $(-1,-1)$ curves that are transversal to the $\IP^2$. 
The result is that at degree $3$ hyperplane class we obtain the following contribution to the free energy from $2$ crosscap configurations
\eqn\degftc{
{\cal F}^{0,2}_3=q_3^2-q_1q_3^2-q_2q_3^2+q_1q_2q_3^2.
}
This is in agreement with the Chern-Simons theory result presented in appendix B.

\newsec{Topological Vertex Computation}

Using large $N$ duality, it was recently proposed \AKMV\ that the free energy of
closed topological strings on a toric
manifold can be computed using a cubic field theory, namely a topological vertex
and gluing rules. In this section we present
a prescription to compute all genus topological string amplitudes
on orientifolds with an external
``${\IR\IP}^2$ leg" by using the topological vertex formalism. We will also explicitly
show that this prescription is equivalent to the large $N$ dual Chern-Simons computation.

\subsec{General prescription}

\ifig\vertexfig{Toric diagram for the quotient $X/I$ of a local, toric Calabi-Yau manifold
with a single ${\IR\IP}^2$.}
{\epsfxsize2in\epsfbox{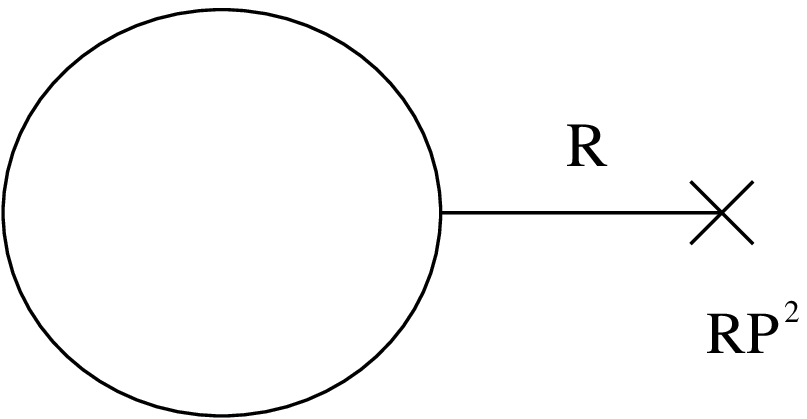}}
Consider a quotient $X/I$ of a local, toric Calabi-Yau manifold $X$ by an involution $I$
which can be represented as in \vertexfig. We have a
bulk geometry, represented by the blob, attached to an ${\IR\IP}^2$ through an
edge associated to the representation $R$. Let us denote by
${\cal O}_{R}$ the amplitude
for the blob with the external leg. We propose
the following formula for total partition function:
\eqn\vertextotal{
Z=\sum_{R=R^T} \CO_R  Q^{\ell(R)/2}
(-1)^{{1\over 2}(\ell(R)\mp r(R))}
}
where the sum is over all self-conjugate
representations $R$. Here $r(R)$ denotes the rank of $R$, as
in \qdSONUN, and $Q=e^{-s}$ is the exponentiated
K\"ahler parameter corresponding to the ${\IR\IP}^2$. The $\mp$ sign is correlated with the
choice of $\pm$ sign for the crosscaps, and corresponds to the $SO/Sp$ gauge duals,
respectively.

The prescription \vertextotal\ comes from the action of the involution $I$ on the partition function
on the covering space, which is given by
\eqn\covering{
Z=\sum_{R} \CO_R (t_i) \CO_{R^T} (t_i) Q^{\ell(R)}
(-1)^{\ell(R)}.
}
where the K\"ahler parameters have been identified in the way prescribed by the involution. 
The involution $I$ maps one half of the toric diagram onto the other half, reversing the orientation of the middle leg. 
The resulting partition function is the one given by \vertextotal.

We are presently investigating in more details the origin of \vertextotal. Having a clear understanding of this formula will 
probably allow us to define a similar prescription for involutions with a fixed locus, like the $I^+$ of \AAHV.

The restriction to self-conjugate representations may appear surprising at first sight. But in the topological vertex formalism, inverting the orientation of one edge sends $R$ to its transpose $R^T$ (and also introduces a factor of $(-1)^{\ell (R)}$). Therefore, since the ${\IR\IP}^2$ leg is unoriented, its partition function must sum only over self-conjugate representations, which are the only representations consistent with the involution $I$.

It is interesting to note that the formula \vertextotal\ is very similar to the formula for quantum dimensions of $SO/Sp$ gauge
group in terms of $U(N)$ quantum dimensions \qdSONUN. Both formulas share the constraint $R=R^T$ and the factor of
$(-1)^{{1\over 2}(\ell(R)\mp r(R))}$. This gives a geometrical argument, from the topological vertex formalism, for the appearance of $SO/Sp$ gauge groups on the Chern-Simons side.

\subsec{Examples}

We now consider two examples of the above prescription.

1) {\it Orientifold of the resolved conifold}. The simplest example is the orientifold of the resolved conifold first considered in
\SV, which we reviewed in section 3.
In that case, the toric diagram is very simple and has been drawn in the left hand side of 
\orienticon. The rule \vertextotal\ gives:
\eqn\firstex{
F= -\log \biggl\{ \sum_{R=R^T}   C_{\cdot \cdot R} Q^{\ell(R)/2}
(-1)^{{1\over 2}(\ell(R)\mp r(R))} \biggr\} .}
This should equal the free energy of Chern-Simons on the sphere for the gauge groups
$SO/Sp$ \sopcs. The free energy
\firstex\ can be indeed computed exactly by using the following
key formula due to Littlewood \refs{\Li,\Macdonald}:
\eqn\little{
\sum_{R=R^T} s_R(x_i)(-1)^{{1\over 2}(\ell(R)\mp r(R))} = \prod_{i=1}^{\infty}(1\pm x_i)
\prod_{1\le i<j<\infty} (1-x_i x_j).}
Since $C_{R \cdot \cdot}=W_R=s_R(q^{-i+{1\over 2}})$, we can compute \firstex\ by
setting $x_i=q^{-i+{1\over 2}}Q^{1\over 2}$ in the r.h.s. of \little. First of all, notice that
\eqn\unpart{
\prod_{i,j}(1-q^{-i-j+1}Q)=\exp \biggl\{ -\sum_{n=1}^{\infty}
{Q^n \over n(q^{n\over 2} -q^{-{n\over 2}})^2} \biggr\}.
}
Also, we can easily compute that
\eqn\unpart{
\prod_{i}(1\mp q^{-i+{1\over 2}}Q^{1\over 2})=\exp \biggl\{ -\sum_{n=1}^{\infty}
{(\pm 1)^n Q^{n\over 2} \over n(q^{n\over 2} -q^{-{n\over 2}})} \biggr\},
}
and from this we easily check that, indeed, the free energy computed in
\firstex\ equals \sopcs.

2) {\it Local ${\IP}^2$ attached to ${\IR\IP}^2$}.
The second example to consider is local ${\IP}^2$ attached to a single
${\IR\IP}^2$, whose toric diagram is drawn in \ptworptwo, and which was
discussed before from the point of view of
geometric transitions. The amplitude for this geometry is given by
\vertextotal\ with
\eqn\orrp{
{\cal O}_{R}= \sum_{R_i} q^{\sum_i \kappa_{R_i}} (-1)^{\sum_i \ell(R_i)}
 C_{\cdot R_3 R_1^T} C_{\cdot R_2 R_3^T} C_{R_1 R_2^T R}e^{-t \sum_i \ell(R_i)},
}
and $t$ is the K\"ahler parameter of local ${\IP}^2$. If we now compare
this expression to the one obtained by geometric transition in \prtwo, we find that
both amplitudes are equal if
\eqn\magic{
{1\over S_{00}^{SO(N)/Sp(N)}}\sum_{R=R^T}   C_{R_1 R_2^T R} Q^{\ell(R)/2}
(-1)^{{1\over 2}(\ell(R)\mp r(R))} = q^{-{\kappa_{R_2}\over 2}}
Q^{{1\over 2}(\ell(R_1) + \ell (R_2))}
{\cal W}^{SO(N)/Sp(N)}_{R_1 R_2},}
where we have taken into account that the partition function of the geometry in \ptworptwo\ also
includes a $t$-independent piece which equals $S_{00}^{SO(N)/Sp(N)}$.
The r.h.s. of \magic\ involves the Hopf link invariant for the gauge groups $SO/Sp$,
where we put $\lambda=Q^{-1}$. Notice that $Q^{{1\over 2}(\ell(R_1) + \ell (R_2))}
{\cal W}^{SO(N)/Sp(N)}_{R_1 R_2}$ is a polynomial in $Q^{1\over 2}$, while
the l.h.s. of \magic\ is {\it a priori} an infinite series in $Q^{1\over 2}$.
The identity \magic\ can be easily proven in the simple case that $R_1$ (or $R_2$) is the trivial
representation, by using again the key identity \little. Let us sketch the proof
in the $Sp$ case, the $SO$ case being similar.

First, notice that, as we have just shown in the example of the orientifold of the conifold,
$S_{00}^{SO(N)/Sp(N)}$ equals the l.h.s. of \little\ evaluated at
$x_i=q^{-i +{1\over 2}}Q^{1\over 2}$. Let us consider the l.h.s. of \little\ for the
$+$ sign (i.e. the $-$ sign in the r.h.s.), but now
evaluated at $x_i=q^{\mu_i -i+{1\over 2}}Q^{1\over 2}$, where $\mu=\{ \mu_i\}_{i=1,\cdots, d(\mu)}$
is the partition corresponding to the representation $R_1^T$. After dividing it by
$S_{00}^{Sp(N)}$ we find a finite product
\eqn\quot{
 \prod_{i=1}^{d(\mu)} {1 - q^{\mu_i -i+{1\over2}} Q^{1\over 2}
\over 1 - q^{-i+{1\over2}} Q^{1\over 2}} \prod_{1\le i <j \le d(\mu)}
{1 - q^{\mu_i +\mu_j-i-j+1} Q^{1\over 2}
\over 1 - q^{-i-j+1} Q^{1\over 2}}
\prod_{i=1}^{d(\mu)}\prod_{v=1}^{\mu_i} (1-q^{\mu_i -i -d(\mu) -v +1}).}
After some massaging, and using the explicit formulae \qdUN, \qdSON, as well
as the relation \SOSpreln, it is easy to see that \quot\ equals
\eqn\quotres{
Q^{\ell(\mu)\over 2} {\CW_{R_1}^{Sp(N)}(\lambda=Q^{-1})\over W_{R_1}},
}
where $W_R =s_R(q^{-i+{1\over 2}})$. Using now that $C_{R R_1\cdot}=W_{R R_1^T}
q^{\kappa_{R_1}/2}$
as well as the explicit formula \hopfschur, we see that indeed \magic\ is satisfied
when $R_2=\cdot$ in the case of $Sp(N)$.

Although we don't have a full proof of \magic\ in general, we have
checked it in many cases. This shows indeed that the topological vertex calculation
and the geometric
transition computation give the same result for this geometry, and indeed for all
the geometries
of the form depicted in \vertexfig.

\newsec{Discussion and Open Problems}

In this paper we have seen how to compute topological string amplitudes on 
a certain class of Calabi-Yau orientifolds, by using geometric transitions involving 
$SO/Sp$ Chern-Simons theory, the topological vertex formalism, and localization techniques. 
This allows us to extract BPS invariants counting higher genus curves with one and two crosscaps. 

This work can be extended in various ways. First of all, it would be very interesting to 
consider Calabi-Yau orientifolds in which the involution has fixed loci, like for example 
the ones considered in \AAHV. In this case, the geometric transition of \SV\ is no longer 
useful and one has to find other ways of implementing a Chern-Simons dual description. In the 
context of the topological vertex formalism, we should find the right prescription to deal with 
fixed point loci, by using perhaps the group-theoretic results of \Li\ 
for $SO/Sp$. Secondly, one should consider open string amplitudes by adding 
Lagrangian D-branes, and to clarify in this way the BPS content of $SO/Sp$ Chern-Simons invariants 
of knots and links. 

It would be also very important to clarify some issues that appeared in the 
orientifolds that we studied here. For example, one would like to have a more detailed 
derivation of the multicovering formulae for amplitudes involving two crosscaps and of the 
choice of annulus operator we made, as well as a more rigorous justification of the localization 
techniques we used. We expect to report on these problems in future work.  
\bigskip
{\it Acknowledgments.} 
We would like to thank S. Kadir for collaboration at the initial stages of this project and
I. Antoniadis, D.-E. Diaconescu and N. Wyllard for useful conversations. The work of V.B. was supported by a Rhodes 
scholarship. The work of B.F. was supported in part by DOE grant DE-FG02-96ER40959.

\appendix{A}{Subsets of Young tableaux}

To compute the tensor product decomposition of irreducible representations of $SO(N)$ and $Sp(N)$ using Littlewood's technique as 
explained in \characSON, we had to use four different subsets of Young tableaux: $\{ \delta \}$ and $\{ \gamma \}$ for $SO(N)$, and 
$\{ \beta \}$ and $\{ \alpha \}$ for $Sp(N)$. These four sets are defined as follows \Li.

$\{\delta \}$ is the set of all partitions into even parts only: $\{ \tableau{2}, \tableau{4}, \tableau{2 2}, ... \}$.

$\{\beta \}$ is the set of all partitions such that there are an even number of parts of any given magnitude: $\{\tableau{1 1}, \tableau{2 2}, 
\tableau{1 1 1 1}, ... \}$.

To define the two remaining sets we have to use the Frobenius notation \refs{\Li, \Ra}. In this notation, a Young tableau is described by an 
array of pair of numbers. The number of pairs is equal to the number of boxes in the leading diagonal of the tableau; the upper number of the 
pair is the number of boxes to the right and the lower number is the number of boxes below. For example, the Young tableau $\tableau{3 3 1}$ 
is described in the Frobenius notation by $\pmatrix{ 2 & 1 \cr 2 & 0 \cr}$.

Using this notation we can define the two remaining sets. Consider Young tableaux defined in the Frobenius notation by
\eqn\frobgen{
\pmatrix{ a_1 & a_2 & a_3 & a_4 & ... \cr b_1 & b_2 & b_3 & b_4 & ... \cr}
.}

$\{ \gamma \}$ is the set of Young tableaux such that $a_i = b_i+1~~ \forall~~ i$: $\{ \tableau{2}, \tableau{3 1}, \tableau{4 1 1},... \}$.

$\{ \alpha \}$ is the set of Young tableaux such that $a_i+1 = b_i~~ \forall~~ i$: $\{ \tableau{1 1}, \tableau{2 1 1}, \tableau{3 1 1 1},...\}$.

Note that $\{ \beta \}$ and $\{ \alpha \}$ are respectively related to $\{ \delta \}$ and $\{ \gamma \}$ by taking the
transpose of the representations, where by transpose we mean exchanging rows and columns.

\appendix{B}{Results in the General Case}

Here we present the results for the full generating functionals given by \functionalsi. The $\pm$ sign corresponds to $Sp$ and $SO$, 
respectively. Of course, the oriented contribution for $q_3=0$ agrees with previous results for the local del Pezzo $dP_3$ with one 
K\"ahler parameter sent to infinity \refs{\CKYZ,\AMV}, and if we set $q_{1,2}=0$ we recover the results presented in Tables $1-16$ (taking 
into account the $1/2$ factor in the definition of the $c=0$ generating functional).

We computed the results up to degree 5 in $e^{-t}$, but we will present only the results up to degree 3 as the higher degree results 
are rather cumbersome.

\eqn\fullresulti{\eqalign{
\CF_0^{0,0}=&~q_1+q_2+{1 \over 2}q_3,\cr
\CF_0^{0,1}=& \pm [q_3^{1/2}],\cr
\CF_0^{0,2}=&~0,\cr
\CF_1^{0,0}=&~3-2(q_1+q_2+q_3)+(q_1q_2+q_2q_3+q_1q_3),\cr
\CF_1^{0,1}=&\pm [-2q_3^{1/2}+( q_1 q_3^{1/2}+q_2q_3^{1/2})],\cr
\CF_1^{0,2}=&~0,
}}

\eqn\fullresultii{\eqalign{
\CF_2^{0,0}=&-6+5(q_1+q_2)+7q_3-4q_1q_2-6(q_1q_3+q_2q_3)+4q_1q_2q_3+{1 \over 2} (q_1^2 q_3 + q_2^2 q_3)\cr
&-q_3^2+(q_1 q_3^2+q_2 q_3^2) - q_1 q_2 q_3^2,\cr
\CF_2^{0,1}=&\pm[5 q_3^{1/2} -4(q_1q_3^{1/2}+q_2q_3^{1/2}) +3 q_1 q_2 q_3^{1/2} -3q_3^{3/2}+2(q_1 q_3^{3/2} + q_2 q_3^{3/2}) - q_1 q_2 q_3^{3/2}],
\cr
\CF_2^{0,2}=&~0,\cr
\CF_3^{0,0}=&~27 - 32(q_1 +q_2) - 42q_3 + 35 q_1 q_2 + 48(q_1 q_3 + q_2 q_3)-50 q_1 q_2 q_3 + 7(q_1^2 + q_2^2)\cr
&+ 15 q_3^2 - 6(q_1^2 q_2 + q_1 q_2^2)- 10(q_1^2 q_3 + q_2^2 q_3)-16(q_1 q_3^2 + q_2 q_3^2)+8(q_1^2 q_2 q_3 + q_1 q_2^2 q_3) \cr
&+ 3 (q_1^2 q_3^2+q_2^2 q_3^2) - 2(q_1^2 q_2 q_3^2 + q_1 q_2^2 q_3^2) + 15 q_1 q_2 q_3^2,\cr
\CF_3^{0,1}=&\pm[-32 q_3^{1/2} +35(q_1 q_3^{1/2} + q_2 q_3^{1/2})-36q_1q_2q_3^{1/2}-6(q_1^2 q_3^{1/2} + q_2^2 q_3^{1/2})\cr
&+(q_1^2 q_2 q_3^{1/2}+q_1 q_2^2 q_3^{1/2}) +30 q_3^{3/2}-30(q_1 q_3^{3/2} + q_2 q_3^{3/2})+4(q_1^2 q_3^{3/2} + q_2^2 q_3^{3/2})\cr
& +28 q_1 q_2 q_3^{3/2}-3(q_1^2 q_2 q_3^{3/2} + q_1 q_2^2 q_3^{3/2}) -4 q_3^{5/2} +3(q_1 q_3^{5/2}+ q_2 q_3^{5/2})-2 q_1 q_2 q_3^{5/2}],\cr
\CF_3^{0,2}=&~q_3^2 -(q_1 q_3^2 +q_2 q_3^2) + q_1 q_2 q_3^2,\cr
\CF_3^{1,0}=&~10-9 (q_1+q_2+q_3)+8(q_1q_2 +q_1q_3 +q_2q_3)-7q_1q_2q_3,\cr
\CF_3^{1,1}=&\pm[-9 q_3^{1/2} +8 (q_1 q_3^{1/2} + q_2 q_3^{1/2}) - 7 q_1 q_2 q_3^{1/2}+7 q_3^{3/2} - 6(q_1 q_3^{3/2} + q_2 q_3^{3/2}) +5 q_1 q_2 q_3^{3/2}],\cr
\CF_3^{1,2}=&~0.
}}

\listrefs

\end